\documentclass[12pt,letterpaper]{article}
\pdfoutput=1
\usepackage{jheppub}
\usepackage{amsfonts, amsthm}
\usepackage{amsmath}
\usepackage{amssymb}
\usepackage{txfonts}
\usepackage{graphicx}
\usepackage{mathrsfs}
\usepackage{nicefrac}
\usepackage{array}
\usepackage[english]{babel}
\usepackage[utf8]{inputenc}
\usepackage{slashed}
\hypersetup{unicode}

\newcommand{\eq}{\begin{equation}}
\newcommand{\feq}{\end{equation}}
\newcommand{\eqn}{\begin{eqnarray}}
\newcommand{\feqn}{\end{eqnarray}}
\font\mybb=msbm10 at 12pt
\def\bb#1{\hbox{\mybb#1}}

\def\bR {\bb{R}}

\newtheorem{prop}{Proposition}
\DeclareMathAlphabet{\mathpzc}{OT1}{pzc}{m}{it}

\title{Fluid dynamics on ultrastatic spacetimes and dual black holes}

\author{Dietmar Klemm$^{a,b}$}
\author{and Andrea Maiorana$^a$}
\affiliation{$^a$ Dipartimento di Fisica, Universit\`a di Milano, \\
\hspace*{0.15cm} Via Celoria 16, 20133 Milano, Italy. \\
$^b$ INFN, Sezione di Milano, \\
\hspace*{0.15cm} Via Celoria 16, 20133 Milano, Italy.
}
\emailAdd{dietmar.klemm@mi.infn.it}
\emailAdd{aerdna.anaroiam@gmail.com}
\preprint{IFUM-1023-FT}

\abstract{We show that the classification of shearless and incompressible stationary fluid flows on
ultrastatic manifolds is equivalent to classifying the isometries of the spatial sections $(\Sigma,\bar g)$.
For a flow on the closed Einstein static universe $\mathbb{R}\times\text{S}^2$ this leaves only
one possibility, since on the 2-sphere all Killing fields are conjugate to each other, and it is well-known
that the gravity dual of such a (conformal) fluid is the spherical Kerr-Newman-AdS$_4$ black hole.
On the other hand, in the open Einstein static universe $\mathbb{R}\times\text{H}^2$ the situation
is more complicated, since the isometry group $\text{SL}(2,\mathbb{R})$ of $\text{H}^2$
admits elliptic, parabolic and hyperbolic elements. One might thus ask what the gravity duals
of the flows corresponding to these three different cases are. Answering this question is one of the
scopes of this paper. In particular we identify the black hole dual to a fluid that is purely translating
on the hyperbolic plane. Although this lies within the Carter-Pleba\'nski class, it has never been
studied in the literature before, and represents thus in principle a new black hole solution in AdS$_4$.
For a rigidly rotating fluid in $\mathbb{R}\times\text{H}^2$ (holographically dual to the hyperbolic
KNAdS$_4$ solution), there is a certain radius where the velocity reaches the speed of light, and thus
the fluid can cover only the region within this radius. Quite remarkably, it turns out that the boundary
of the hyperbolic KNAdS$_4$ black hole is conformal to exactly that part of $\mathbb{R}\times\text{H}^2$
in which the fluid velocity does not exceed the speed of light. Thus, the correspondence between AdS
gravity and hydrodynamics automatically eliminates the unphysical region.
We extend these results to establish a precise mapping between possible flows on ultrastatic
spacetimes (with constant curvature spatial sections) and the parameter space of the Carter-Pleba\'nski
solution to Einstein-Maxwell-AdS gravity.
Finally, we show that the alternative description of the hyperbolic KNAdS$_4$ black hole in
terms of fluid mechanics on $\mathbb{R}\times\text{S}^2$ or on flat space (both conformal to the
open Einstein static universe) is dynamical and consists of a contracting or expanding vortex.
}


\begin{document}
\maketitle
\flushbottom

\section{Introduction}

The AdS/CFT correspondence has provided us with a powerful tool to get insight into the dynamics of
certain field theories at strong coupling by studying classical gravity solutions. In the long wavelength
limit, where the mean free path is much smaller then any other scale,
one expects that these interacting field theories admit an effective hydrodynamical description.
In fact, it was shown in \cite{Bhattacharyya:2008jc}\footnote{Analogous results in four and higher
dimensions were obtained in \cite{VanRaamsdonk:2008fp}
and \cite{Bhattacharyya:2008mz,Haack:2008cp} respectively.}
that the five-dimensional Einstein equations with negative
cosmological constant reduce to the Navier-Stokes equations on the conformal boundary of AdS$_5$.
The analysis of \cite{Bhattacharyya:2008jc} is perturbative in a boundary derivative expansion,
in which the zeroth order terms describe a conformal perfect fluid. The coefficient of the first
subleading term yields the shear viscosity $\eta$ and confirms the famous result $\eta/s=1/(4\pi)$
by Policastro, Son and Starinets \cite{Policastro:2001yc}, which was obtained by different methods.
Subsequently, the correspondence between AdS gravity and fluid dynamics (cf.~\cite{Rangamani:2009xk}
for a review) was extended in
various directions, for instance to include forcing terms coming from a dilaton \cite{Bhattacharyya:2008ji}
or from electromagnetic fields (magnetohydrodynamics) \cite{Hansen:2008tq,Caldarelli:2008ze}.
The gravitational dual of non-relativistic incompressible fluid flows was obtained
in \cite{Bhattacharyya:2008kq}.

In addition to providing new insights into the dynamics of gravity, the map between hydrodynamics
and AdS gravity has contributed to a better understanding of various issues in fluid dynamics.
One such example is the role of quantum anomalies in hydrodynamical transport \cite{Son:2009tf}.
Moreover, it has revealed beautiful and unexpected relationships between apparently very
different areas of physics, for instance it was argued in \cite{Caldarelli:2008mv} that the Rayleigh-Plateau
instability in a fluid tube is the holographic dual of the Gregory-Laflamme instability of a black
string\footnote{Note in this context that the instability of the effective fluid that describes
higher-dimensional asymptotically flat black branes, analyzed in \cite{Camps:2010br}, is not of
the Rayleigh-Plateau type, but rather one in the sound modes.}.
The hope is that eventually the fluid/gravity correspondence may shed light on fundamental
problems in hydrodynamics like turbulence. Another possible application is the quark-gluon plasma
created in heavy ion collisions, where perturbative QCD does not work, and lattice QCD
struggles with dynamic situations, cf.~e.g.~\cite{Myers:2008fv}.
We will come back to this point in section \ref{fin-rem}.

Here we will use fluid dynamics to make predictions on which types of black holes can
exist in four-dimensional Einstein-Maxwell-AdS gravity. In particular, we shall classify all possible
stationary equilibrium flows on ultrastatic manifolds with constant curvature spatial sections, and
then use these results to predict (and explicitely construct) new black hole solutions.

The remainder of this paper is organized as follows: In the next section, we briefly review the basics
of conformal hydrodynamics. In section \ref{stat flow ultrastatic st} we consider shearless and
incompressible stationary fluids on ultrastatic manifolds, and show that the classification of such
flows is equivalent to classifying the isometries of the spatial sections
$(\Sigma,\bar g)$\footnote{Since every static spacetime is conformally ultrastatic, these results extend
of course to arbitrary static spacetimes in the case of conformal hydrodynamics.}. This is then
applied to the three-dimensional case with constant curvature spatial sections, i.e., to fluid dynamics
on $\mathbb{R}\times\text{S}^2$, $\mathbb{R}\times\text{H}^2$ and Minkowski space
$\mathbb{R}\times\text{E}^2$.
It is shown that, up to isometries, the flow on the 2-sphere is unique, while there are three
non-conjugate Killing fields on the hyperbolic plane and two on the Euclidean plane.
In almost all cases, it turns out that the fluid can cover only a part of the manifold, since
there exist regions where the fluid velocity exceeds the speed of light. This property is quite
obvious for rigid rotations on $\text{H}^2$ or $\text{E}^2$: Here there is a certain radius where the
velocity reaches the speed of light, and thus the fluid can cover only the region within this radius.
Due to the diverging gamma factor at the boundary of the fluid, the global thermodynamic variables
like energy, angular momentum, entropy and electric charge are infinite in these cases.
Nevertheless, we show that a local form of the first law of thermodynamics still holds.
At the end of section \ref{stat flow ultrastatic st}, we transform the rigidly rotating conformal fluid
on the open Einstein static universe to $\mathbb{R}\times\text{S}^2$ and to Minkowski space
(this is possible since both are conformal to $\mathbb{R}\times\text{H}^2$), and shew that this
yields contracting or expanding vortex configurations.

In section \ref{dual-AdS-BH}, the gravity duals of the hydrodynamic flows considered in
\ref{stat flow ultrastatic st} are identified. Although they all lie within the Carter-Pleba\'nski
class \cite{Carter:1968ks,Plebanski:1975}, many of them have never been studied in the literature
before, and represent thus in principle new black hole solutions in AdS$_4$.
Quite remarkably, it turns out that the boundary of these black holes are conformal to
exactly that part of $\mathbb{R}\times\text{S}^2$, $\mathbb{R}\times\text{H}^2$ or Minkowski
space in which the fluid velocity does not exceed the speed of light. Thus, the correspondence between
AdS gravity and hydrodynamics automatically eliminates the unphysical region.

We conclude in section \ref{fin-rem} with some final remarks. In appendix \ref{app-CP}, our results
are extended to establish a precise mapping between possible flows on ultrastatic
spacetimes (with constant curvature spatial sections) and the parameter space of the Carter-Pleba\'nski
solution to Einstein-Maxwell-AdS gravity.
The proofs of some propositions are relegated to appendix \ref{app-proof}.

Note that a related, but slightly different approach was adopted in \cite{Mukhopadhyay:2013gja}, where
uncharged fluids in Papapetrou-Randers geometries were considered. In these flows, the fluid
velocity coincides with the timelike Killing vector of the spacetime (hence the fluid is at rest in this frame),
and the Cotton-York tensor has the form of a perfect fluid (so-called `perfect Cotton geometries').
We will see below that there is some overlap between the bulk geometries dual to such flows,
constructed explicitely in \cite{Mukhopadhyay:2013gja}, and the solutions obtained here.

Throughout this paper we use calligraphic letters ${\cal T},{\cal V},{\cal S},\ldots$ to indicate
local thermodynamic quantities, whereas $T,V,S,\ldots$ refer to the whole fluid configuration.
$\mu$ and $\phi_{\text e}$ are local and global electric potentials respectively.

\newpage

\section{Conformal hydrodynamics}
\label{conf-hydro}

Consider a charged fluid on a $d$-dimensional spacetime. The equations of hydrodynamics are simply
the conservations laws for the stress tensor $T^{\mu\nu}$ and the charge current $J^{\mu}$,
\begin{equation}
\nabla_\mu T^{\mu\nu}=0\,, \qquad \nabla_\mu J^{\mu} = 0\,.
\end{equation}
Since fluid mechanics is an effective description at long distances, valid when the fluid variables
vary on scales much larger than the mean free path, it is natural to expand the energy-momentum
tensor, charge current and entropy current $J^{\mu}_S$ in powers of derivatives. At zeroth order in
this expansion, one has the perfect fluid form \cite{Andersson:2006nr}
\begin{equation}
T^{\mu\nu}_{\text{perf}} = (\rho+{\cal P})u^\mu u^\nu+{\cal P} g^{\mu\nu}\,, \qquad J^\mu_{\text{perf}} =
\rho_{\text e}u^\mu\,, \qquad {J^\mu_S}_{\text{perf}} = s u^\mu\,,
\end{equation}
where $u$ denotes the velocity profile, and $\rho$, ${\cal P}$, $\rho_{\text e}$ and $s$ are the
energy density, pressure, charge density and entropy density respectively, measured in the local rest
frame of the fluid.

At first subleading order, one obtains the dissipative contributions \cite{Andersson:2006nr}
\begin{equation} \label{T-diss}
T^{\mu\nu}_{\text{diss}}=-\zeta\vartheta P^{\mu\nu}-2\eta\sigma^{\mu\nu}+(q^\mu u^\nu+
q^\nu u^\mu)\,, \qquad J^\mu_{\text{diss}} = q^\mu_{\text e}\,, \qquad {J^\mu_S}_{\text{diss}} =
\frac{q^\mu - \mu q^\mu_{\text e}}{\cal T}\,,
\end{equation}
where
\begin{equation}
P^{\mu\nu}=g^{\mu\nu}+u^\mu u^\nu\,,
\end{equation}
and
\begin{equation} \label{a-theta-sigma}
a^\mu=u^\nu\nabla_\nu u^\mu\,, \qquad \vartheta=\nabla_\mu u^\mu\,, \qquad
\sigma^{\mu\nu}=\frac{1}{2}(P^{\mu\rho}\nabla_\rho u^\nu+P^{\nu\rho}\nabla_\rho u^\mu)-\frac{1}{d-1}\vartheta P^{\mu\nu}\,,
\end{equation}
\begin{equation}
q^\mu=-\kappa P^{\mu\nu} (\partial_\nu + a_\nu){\cal T}\,, \qquad
q_{\text e}^\mu = -D P^{\mu\nu}\partial_\nu\frac{\mu}{\cal T} \label{subl-currents}
\end{equation}
denote the acceleration, expansion, shear tensor, heat flux and diffusion current respectively.
Moreover, ${\cal T}$ and $\mu$ are the local temperature and electric potential, $\zeta$
is the bulk viscosity, $\eta$ the shear viscosity,
$\kappa$ the thermal conductivity and $D$ the diffusion coefficient. Note that the equations
\eqref{subl-currents} are the relativistic generalizations of Fourier's law of heat condution and
Fick's first law.

At first order in the derivative expansion, the entropy current is no longer conserved, but
obeys \cite{Andersson:2006nr}
\begin{equation} \label{div-JS}
{\cal T}\nabla_\mu J_S^\mu=\frac{q_\mu q^\mu}{\kappa {\cal T}} + \frac{\cal T}{D}
q_{\text e}^\mu q_{\text{e}\mu}+\zeta\vartheta^2+2\eta\sigma_{\mu\nu}\sigma^{\mu\nu}\,.
\end{equation}
If the coefficients satisfy certain non-negativity conditions,
this implies $\nabla_\mu J_S^\mu\geq 0$, and thus entropy is always non-decreasing.
In equilibrium, $ J^\mu_S$ must be conserved, which is the case if and only if
$q^\mu$, $q^\mu_{\text e}$, $\vartheta$ and $\sigma^{\mu\nu}$ all vanish\footnote{In the case
of zero viscosities, $\zeta=\eta=0$, one can in principle allow for nonvanishing expansion and shear
tensor. In particular, for conformal fluids (cf.~below), the bulk viscosity vanishes, and therefore
the third term on the rhs of \eqref{div-JS} is zero without imposing $\vartheta=0$.}.

Since we consider fluids on curved manifolds, we could add to $T^{\mu\nu}$ also terms
constructed from the curvature tensors. In fact, at second order in a derivative
expansion, there is a term proportional to the Weyl tensor of the boundary, cf.~equation (2.10) of
\cite{Bhattacharyya:2008mz}. However, in all explicit examples considered here, the boundary
is three-dimensional, and thus its Weyl tensor vanishes. Note that in three dimensions there
is a possible third order contribution from the Cotton tensor \cite{Mukhopadhyay:2013gja}, but
since our boundary geometries \eqref{conf-bdry-CP} are conformally flat for vanishing
NUT-parameter, this contribution vanishes as well.

In what follows, we specialize to conformal fluids\footnote{For a Weyl-covariant formalism that
simplifies the study of conformal hydrodynamics cf.~\cite{Loganayagam:2008is}.}. Upon a Weyl rescaling
$\tilde{g}_{\mu\nu}=\Omega^2 g_{\mu\nu}$, the energy-momentum tensor must transform as
$\tilde T^{\mu\nu}=\Omega^w T^{\mu\nu}$ for some weight $w$, and hence
\begin{equation}
\tilde{\nabla}_\mu\tilde{T}^{\mu\nu}=\Omega^w\nabla_\mu T^{\mu\nu} + \Omega^{w-1}\left((w+d+2)
T^{\mu\nu}\partial_\mu\Omega - T^\mu{}_\mu\,\partial^\nu\Omega\right)\,,
\end{equation}
from which we learn that $w=-(d+2)$ and $T^\mu{}_\mu=0$ in order for $\tilde{T}$ to be conserved.
The tracelessness of $T$ implies the equation of state $\rho=(d-1){\cal P}$ and requires the bulk viscosity
$\zeta$ to be zero. The transformation laws for the fluid variables are
\begin{displaymath}
\tilde{u}=\Omega^{-1}u\,, \quad \tilde{\rho}=\Omega^{-d}\rho\,, \quad \tilde{\cal P}=
\Omega^{-d}{\cal P}\,, \quad \tilde\rho_{\text e} = \Omega^{-(d-1)}\rho_{\text e}\,, \quad
\tilde s = \Omega^{-(d-1)}s\,, \quad \tilde{\cal T} = \Omega^{-1}{\cal T}\,.
\end{displaymath}
Furthermore, the charge- and entropy current transform as
\begin{displaymath}
\tilde{J}^\mu=\Omega^{-d}J^\mu\,, \qquad \tilde{J}_S^\mu=\Omega^{-d}J_S^\mu\,.
\end{displaymath}

Note that, if the charged fluid moves in an external electromagnetic field $F_{\mu\nu}$, its stress
tensor is no more conserved, and the equations of motion become
\begin{equation}
\nabla_\mu T^{\mu\nu}=F^\nu{}_\mu J^\mu\,, \label{eq:MHD}
\end{equation}
where the rhs represents the Lorentz force density. This scenario was studied
in full generality in \cite{Caldarelli:2008ze}. According to the AdS/CFT dictionary, such an external field
is related to the magnetic charge of the dual black hole. Quite surprisingly, it turns out that for all the
magnetically charged black holes considered here, there is no net Lorentz force acting on the dual fluid,
since the electric and magnetic forces exactly cancel. One has thus $F^\nu{}_\mu J^\mu=0$, hence
$T^{\mu\nu}$ is conserved.

At the end of this section, we briefly review the constraints imposed on the thermodynamics
by conformal invariance. First of all, define the grand-canonical potential
\begin{equation}
\Phi = {\cal E} - {\cal T}{\cal S} - \mu{\cal Q}_{\text{e}}\,,
\end{equation}
which satisfies the first law
\begin{equation}
\mathrm{d}\Phi = -{\cal S}\mathrm{d}{\cal T} - {\cal P}\mathrm{d}{\cal V} -
{\cal Q}_{\text e}\mathrm{d}\mu\,. \label{first-law}
\end{equation}
Conformal invariance and extensivity imply that $\Phi$ must have the
form \cite{Bhattacharyya:2007vs}
\begin{equation}
\Phi = -{\cal V}{\cal T}^d h(\psi)\,,
\end{equation}
for some function $h(\psi)$, where $\psi:=\mu/{\cal T}$. The remaining thermodynamic
quantities are then easily obtained using \eqref{first-law},
\begin{equation} \label{remaining-h}
{\cal P} = \frac{\rho}{d-1} = {\cal T}^d h(\psi)\,, \quad \rho_{\text e} = \frac{{\cal Q}_{\text e}}{\cal V}
= {\cal T}^{d-1} h'(\psi)\,, \quad s = \frac{\cal S}{\cal V} = {\cal T}^{d-1}(dh(\psi) - \psi h'(\psi))\,.
\end{equation}

\section{Equilibrium flows in ultrastatic spacetimes}\label{stat flow ultrastatic st}

We will now focus on conformal fluids in ultrastatic spacetimes, and explain how the equilibrium flows
can be classified using the isometries of the spatial sections.

A $d$-dimensional spacetime $(M,g)$ is said to be \textit{ultrastatic} if there are a timelike Killing field
$\xi$ such that $\xi_\mu\xi^\mu=-1$ and a hypersurface $\Sigma$ orthogonal to $\xi$. In such a
spacetime one can always choose a coordinate system such that
\begin{equation}
g=-\mathrm{d}t^2+\bar{g}_{ij}\mathrm{d}x^i \mathrm{d}x^j\,,
\end{equation}
where $\bar{g}$ is the induced metric on $\Sigma$. The velocity field $u$ for a flow on $M$
can be written as
\begin{equation} \label{u=gamma(1,v)}
u^\mu=\gamma(1,v^i)\,,
\end{equation}
and the constraint $u_\mu u^\mu=-1$ implies that $\gamma^2=1/(1-v^2)$, where
$v^2:=\bar{g}_{ij}v^i v^j$.
We assume that the fluid is stationary in the frame $(t,\vec{x})$, that is $\partial_t u^\mu=0$.
Equ.~\eqref{u=gamma(1,v)} defines then a vector field $v$ on $\Sigma$.
Note that the property of ultrastaticity is not conserved under general Weyl rescalings. Thus when
we say that a conformal spacetime $(M,[g])$ is ultrastatic, we mean that it has \textit{some} metric representative $g$ which is ultrastatic.

As was explained in section \ref{conf-hydro}, a fluid is in equilibrium when the entropy current is
conserved, which implies that the flow must be shearless and incompressible. The classification
of such flows becomes quite easy if we use the following proposition, proven in appendix \ref{app-proof}.

\begin{prop} \label{prop-killing}
$\sigma^{\mu\nu}=0$ and $\vartheta=0$ $\Leftrightarrow$ $v$ is a Killing field for $(\Sigma,\bar{g})$.
\end{prop}

The classification of shearless and incompressible flows on ultrastatic manifolds is thus equivalent
to classifying the isometries of the spatial sections $(\Sigma,\bar g)$.

In equilibrium, the dissipative contribution to $T_{\mu\nu}$ in
\eqref{T-diss} vanishes, and the stress tensor is just
\begin{equation}\label{d-dim conf fl stress tens}
T^{\mu\nu} = {\cal P}(d\, u^\mu u^\nu+g^{\mu\nu})\,,
\end{equation}
where we used the equation of state $\rho=(d-1){\cal P}$ of conformal fluids. The solution
of the Navier-Stokes equations becomes then particularly simple:

\begin{prop} \label{prop-NS}
When $\sigma^{\mu\nu}=\vartheta=0$ and ${\cal P},u^\mu$ are independent of $t$, the stress tensor
\eqref{d-dim conf fl stress tens} satisfies $\nabla_\mu T^{\mu\nu}=0$ if and only if 
\begin{equation}\label{sol eq NS}
{\cal P}={\cal P}_0\gamma^d
\end{equation}
for some constant ${\cal P}_0$.
\end{prop}

Consider now the heat flux $q^\mu$ and the diffusion current $q_{\text e}^\mu$, given in
\eqref{subl-currents}.

\begin{prop} \label{prop-heat-flux}
If the flow is stationary, incompressible and shearless, then $q^\mu=0$ implies
${\cal T} = \tau\gamma$ for some constant $\tau$.
\end{prop}

\begin{prop} \label{prop-diff-curr}
For a stationary flow, $q_{\text e}^\mu=0$ implies $\mu=\psi{\cal T}$, where $\psi$ is a constant.
\end{prop}

The proofs of propositions \ref{prop-NS}-\ref{prop-diff-curr} are again given in
appendix \ref{app-proof}. With \ref{prop-NS}, \ref{prop-heat-flux} and \ref{prop-diff-curr},
\eqref{remaining-h} becomes
\begin{equation}
{\cal P}_0 = \tau^d h(\psi)\,, \qquad \rho_{\text e} = \tau^{d-1}\gamma^{d-1} h'(\psi)\,, \qquad
s = \tau^{d-1}\gamma^{d-1}(dh(\psi) - \psi h'(\psi))\,. \label{P_0-rho_e-s}
\end{equation}
Finally, the second of these equations implies that the charge current $J^\mu=\rho_{\text e}u^\mu$
is conserved.

At this point, some comments are in order:
We have shown that we can construct all stationary shearless and incompressible fluid configurations
on the spacetime $M$, if we know the Killing fields on the spatial sections $(\Sigma,\bar g)$.
A Killing field $v$ is defined on the whole manifold $\Sigma$, but it gives a physically meaningful flow
only on the subset $U\subset\Sigma$ in which $v^2<1$. Notice that $v^2$ is constant along the
integral curves of $v$, and therefore the flow does not cross the boundary of $U$, where the fluid
moves at the speed of light.
Moreover, we do not need to consider the flow arising from each Killing field $v$, since different flows can
be isometric. Suppose in fact that we have two Killing fields $v,\tilde{v}$ which are related by an isometry $\Psi$, i.e., $\tilde{v}\circ\Psi=\mathrm{d}\Psi\circ v$. In terms of the 1-parameter groups of isometries
$\Phi^{(v)},\Phi^{(\tilde{v})}$ that these fields generate, this condition reads
\begin{equation}\label{Psi-related 1-par groups}
\Phi^{(\tilde{v})}_{\lambda}\circ\Psi=\Psi\circ\Phi^{(v)}_{\lambda}\,.
\end{equation}
When this holds, the flows arising from $v$ and $\tilde{v}$ are physically equivalent.
Now, to the Killing fields $v,\tilde{v}$ correspond two elements $A,\tilde{A}$ in the
Lie algebra $i(\Sigma)$ of the
isometry group $I(\Sigma)$, namely the generators of the 1-parameter subgroups
$\lambda\mapsto\Phi^{(v)}_{\lambda}$ and $\lambda\mapsto\Phi^{(\tilde{v})}_{\lambda}$,
for which equ.~\eqref{Psi-related 1-par groups} becomes
\begin{equation}\label{Psi-related Lie algebra elements}
\tilde{A}=\textup{Ad}_\Psi(A)\,,
\end{equation}
where $\textup{Ad}$ is the adjoint representation of $I(\Sigma)$ on $\mathfrak{i}(\Sigma)$. This reduces
the problem of finding inequivalent flows to the study of the properties of the Lie algebra
$\mathfrak{i}(\Sigma)$ under the adjoint representation.

\subsection{Stationary conformal fluid on the 2-sphere}\label{stat conf fluid sph}

Let us first study the case (partially considered in \cite{Bhattacharyya:2007vs,Caldarelli:2008ze}) in
which the conformal fluid lives on the ultrastatic spacetime
$\mathbb{R}\times\text{S}^2$, with metric given by
\begin{equation}\label{metric sph spacetime}
g=-\mathrm{d}t^2+\ell^2(\mathrm{d}\theta^2+\sin^2\theta\mathrm{d}\varphi^2)\,.
\end{equation}
As was explained above, the 3-velocity of the fluid is $u^\mu=\gamma(1,v^i)$, where $v$ is a Killing
field of $\text{S}^2$. By a rotation, $v$ can be brought to a multiple of any other 
Killing field, say $\partial_\varphi$. Thus we can take $v=\omega\partial_\varphi$, with $\omega\in\mathbb{R}$, without loss of generality. Hence
\begin{equation}
u=\gamma(\partial_t+\omega\partial_\varphi)\,,
\end{equation}
where $\gamma=(1-\omega^2\ell^2\sin^2\theta)^{-1/2}$. This means that the motion of the
fluid in equilibrium is just a rigid rotation on $\text{S}^2$.
The physical constraint $v^2<1$ limits the fluid to polar caps at $|\omega|\ell\sin\theta<1$.
Thus, if we restrict $|\omega|\ell<1$, the physical region $U$ is the whole sphere.

The stress tensor of the fluid is $T^{\mu\nu}=(\rho+{\cal P})u^\mu u^\nu+{\cal P} g^{\mu\nu}$
(dissipative terms vanish because of equilibrium), where $\rho=2{\cal P}$. This gives
\begin{equation}\label{fluid on sphere stress tensor}
T^{\mu\nu}={\cal P}\begin{pmatrix}
3\gamma^2-1 & 0 & 3\gamma^2\omega\\ 
0 & \frac{1}{\ell^2} & 0\\ 
3\gamma^2\omega & 0 & \frac{3\gamma^2-2}{\ell^2\sin^2\theta}
\end{pmatrix}\,,
\end{equation}
which is conserved if
\begin{equation}
{\cal P} = {\cal P}_0\gamma^3\,, \label{P_0gamma^3}
\end{equation}
where we used \eqref{sol eq NS}.
The heat flux $q^\mu$ and diffusion current $q_{\text e}^\mu$ vanish by virtue of propositions
\ref{prop-heat-flux} and \ref{prop-diff-curr}.

We now want to compute the conserved charges associated to the stress tensor $T^{\mu\nu}$ and the
currents $J^\mu$, $J^\mu_S$. These are well-defined only for $|\omega|\ell<1$, since otherwise
the physical region $U$ has a boundary where the Lorentz factor $\gamma$ diverges.
We consider the foliation of spatial surfaces $\Sigma_t\simeq\text{S}^2$ of constant $t$, with induced
metric $\bar g$. In the case $|\omega|\ell<1$ the electric charge and entropy are given respectively by
\begin{equation}
Q_{\text e} = \int_{\Sigma_t} d^2x\sqrt{\bar g}J^t = \frac{4\pi\ell^2\tau^2 h'(\psi)}{1 -
\omega^2\ell^2}\,,
\quad S = \int_{\Sigma_t} d^2x\sqrt{\bar g}J^t_S = \frac{4\pi\ell^2\tau^2(3h(\psi) - \psi h'(\psi))}
{1 - \omega^2\ell^2}\,, \label{Q-S}
\end{equation}
while the total energy $E$ and angular momentum $L$ read
\begin{equation}
E = -\int_{\Sigma_t} d^2x\sqrt{\bar g}{T^t}_\mu\xi^\mu = \frac{8\pi\ell^2\tau^3 h(\psi)}{(1-\omega^2
\ell^2)^2}\,, \label{E}
\end{equation}
\begin{equation}
L = -\int_{\Sigma_t} d^2x\sqrt{\bar g}{T^t}_\mu\chi^\mu = \frac{8\pi\ell^4\tau^3\omega h(\psi)}
{(1-\omega^2\ell^2)^2}\,, \label{L}
\end{equation}
where we used the Killing vectors $\xi=\partial_t$ and $\chi=-\partial_{\varphi}$.
The charges \eqref{Q-S}-\eqref{L} were obtained for the first time in \cite{Bhattacharyya:2007vs}.
The volume $V=4\pi\ell^2$ is fixed and not considered as a thermodynamical variable.
It is straightforward to verify that $E$, $L$, $S$, $Q_{\text e}$, which are functions of the parameters
$\omega,\tau,\psi$, satisfy the first law
\begin{equation}
\mathrm{d}E=\tau\mathrm{d}S+\omega\mathrm{d}L+\tau\psi\mathrm{d}Q_{\text e}\,.
\end{equation}
As a consequence, the intensive variables conjugate to $S,L,Q_{\text e}$ are respectively
\begin{equation}
T = \left(\frac{\partial E}{\partial S}\right)_{L,Q_{\text e}} = \tau\,, \qquad
\Omega = \left(\frac{\partial E}{\partial L}\right)_{S,Q_{\text e}} = \omega\,, \qquad
\phi_{\text e} = \left(\frac{\partial E}{\partial Q_{\text e}}\right)_{S,L} = \tau\psi\,. \label{T-Omega-phi}
\end{equation}
Finally, the grandcanonical potential $G=E-TS-\Omega L-\phi_{\text e}Q_{\text e}$ reads
\begin{equation}
G = -\frac{4\pi\ell^2\tau^3 h(\psi)}{1-\omega^2\ell^2}\,,
\end{equation}
where $\psi=\phi_{\text e}/T$.

\subsection{Stationary conformal fluid on a plane} \label{sec:rot-plane}

We now consider a conformal fluid on three-dimensional Minkowski space
$\mathbb{R}\times\text{E}^2$, with metric
\begin{equation}
g=-\mathrm{d}t^2+\mathrm{d}x^2+\mathrm{d}y^2\,.
\end{equation}
The Killing fields on the plane $\text{E}^2$ are linear combinations of
\begin{equation}
\xi^{(R)} = -y\partial_x+x\partial_y\,, \qquad \xi^{(T_1)} = \partial_x\,, \qquad \xi^{(T_2)} = \partial_y\,.
\end{equation}
By using the commutation relations
\begin{displaymath}
[R,T_1] = T_2\,, \qquad [R,T_2] = -T_1\,, \qquad [T_1,T_2] = 0
\end{displaymath}
of the Euclidean group $\text{ISO}(2)$, it is easy to shew that
\begin{equation} \label{Ad_T(S)}
e^{a\hat{\bf m}\cdot{\bf T}}Re^{-a\hat{\bf m}\cdot{\bf T}} = R + a(m^2T_1 - m^1T_2)\,,
\end{equation}
where $a$ is a constant, $\hat{\bf m}=(m^1,m^2)$ denotes a unit vector, and ${\bf T}=(T_1,T_2)$.
If we choose
\begin{displaymath}
a = \frac{\beta}{\omega}\,, \qquad m^1 = -\frac{\beta^2}{\beta}\,, \qquad m^2 = \frac{\beta^1}{\beta}\,,
\qquad \beta := \sqrt{(\beta^1)^2 + (\beta^2)^2}\,,
\end{displaymath}
\eqref{Ad_T(S)} implies that $\omega R + \beta^1T_1 + \beta^2T_2$ is in the same orbit as $\omega R$
under $\text{ISO}(2)$, as long as $\omega\neq 0$. For $\omega=0$ the spatial fluid velocity is
$v=\beta^1\partial_x + \beta^2\partial_y$, i.e., one has a purely translating fluid
on $\mathbb{R}\times\text{E}^2$, which is dual to a boosted Schwarz\-schild-AdS black hole with flat
horizon. We shall thus assume $\omega\neq 0$ in what follows. In this case, as just explained,
it is (up to isometries) sufficient to consider a fluid that rotates around the origin. If we introduce
polar coordinates $r,\varphi$, the 3-velocity becomes
\begin{equation}
u=\gamma(\partial_t+\omega\partial_\varphi)\,,
\end{equation}
where $\gamma=(1-\omega^2 r^2)^{-1/2}$. Note that the flow is well-defined only for $r<1/\omega$.
At $\omega r=1$ the fluid rotates at the speed of light.

The stress tensor of this configuration is given by
\begin{equation}
T^{\mu\nu} = {\cal P}\begin{pmatrix}
3\gamma^2-1 & 0 & 3\gamma^2\omega\\ 
0 & 1 & 0\\ 
3\gamma^2\omega & 0 & \frac{3\gamma^2-2}{r^2}
\end{pmatrix}\,,
\end{equation}
which is again conserved if \eqref{P_0gamma^3} holds.

\subsection{Stationary conformal fluid on hyperbolic space} \label{sec:fluid-H2}

The last example that we consider is a conformal fluid in equilibrium on $\mathbb{R}\times\text{H}^2$,
with metric given by
\begin{equation} \label{metric hyperb spacetime}
g=-\mathrm{d}t^2+\ell^2(\mathrm{d}\theta^2+\sinh^2\theta\mathrm{d}\varphi^2)\,.
\end{equation}
To begin with a simple scenario, one can just follow what we did for the 2-sphere in subsection
\ref{stat conf fluid sph}, taking the fluid in rigid rotation on the hyperboloid. Most of the results reflect
what we found for the spherical flow. However, there are also some differences: As we shall see,
no matter how small the angular velocity is, there always exists a certain critical distance from the center
of rotation where the fluid moves at the speed of light, and hence the physical region $U$ is always
smaller than the whole hyperboloid $\text{H}^2$. As a consequence, one cannot analyze
the global thermodynamic properties of the system, since the extensive variables diverge. Anyway, we
will show that for this fluid configuration one can make a local thermodynamical analysis to find some
results comparable with those of subsection \ref{stat conf fluid sph}.

While in the spherical case the rigidly rotating flux is the only solution in equilibrium, for the hyperbolic
plane there are different, inequivalent solutions, since this space admits non-conjugate Killing fields.
(The isometry group $\text{SL}(2,\bR)$ of $\text{H}^2$ has parabolic, hyperbolic and elliptic elements). 
We denote the generators of $\text{SL}(2,\bR)$ by $R,B_1,B_2$. These obey
\begin{displaymath}
[R,B_1] = B_2\,, \qquad [R,B_2] = -B_1\,, \qquad [B_1,B_2] = -R\,,
\end{displaymath}
and are represented on the Poincar\'e disk by
\begin{displaymath}
\xi^{(R)} = i(z\partial_z - \bar z\partial_{\bar z})\,, \quad \xi^{(B_1)} = \frac12(1 - z^2)\partial_z
+ \frac12(1 - \bar z^2)\partial_{\bar z}\,, \quad \xi^{(B_2)} = \frac i2(1 + z^2)\partial_z
- \frac i2(1 + \bar z^2)\partial_{\bar z}\,.
\end{displaymath}
The complex coordinate $z$ is related to $\theta,\varphi$ by $z=e^{i\varphi}\tanh\frac{\theta}2$.
One easily shows that
\begin{equation}
e^{\alpha R}(\omega R + \beta B_1)e^{-\alpha R} = \omega R + \beta(B_1\cos\alpha + B_2\sin\alpha)\,,
\end{equation}
and thus a general linear combination $\omega R + \beta^1 B_1 + \beta^2 B_2$ is conjugate to
$\omega R + \beta B_1$, so we can drop $B_2$ without loss of generality. Moreover, one has
\begin{equation}
e^{\chi B_2} R e^{-\chi B_2} = R\cosh\chi + B_1\sinh\chi\,. \label{Ad_B2(R)}
\end{equation}
If $\omega^2>\beta^2$, we can put $\tanh\chi=\beta/\omega$, and \eqref{Ad_B2(R)}
implies that $\omega R+\beta B_1$ is conjugate to $\tilde\omega R$, where
$\tilde\omega:=\omega\sqrt{1-\beta^2/\omega^2}$. This case corresponds to an elliptic
element of $\text{SL}(2,\bR)$, and describes a fluid in rigid rotation on $\text{H}^2$.

For $\omega^2<\beta^2$ (hyperbolic element), use
\begin{displaymath}
e^{\chi B_2} B_1 e^{-\chi B_2} = R\sinh\chi + B_1\cosh\chi\,, \qquad \tanh\chi=\omega/\beta\,,
\end{displaymath}
to show that $\omega R+\beta B_1$ is in the same orbit as $\tilde\beta B_1$, with
$\tilde\beta:=\beta\sqrt{1-\omega^2/\beta^2}$.

Finally, for $\omega^2=\beta^2$ (parabolic element), one can set $\omega=\beta$ without loss of
generality, since the case $\omega=-\beta$ is related to this by the discrete isometry $J$ obeying
\begin{displaymath}
JRJ^{-1} = -R\,, \qquad JB_1J^{-1} = B_1\,, \qquad JB_2J^{-1} = -B_2\,.
\end{displaymath}
In the complex coordinates $z,\bar z$, the transformation $J$ acts as $z\to\bar z$.
As representative in this last case we can thus take the Killing vector $\omega(\xi^{(R)} + \xi^{(B_1)})$.
Notice that due to
\begin{equation}
e^{\chi B_2}(R+B_1)e^{-\chi B_2} = e^{\chi}(R+B_1)\,,
\end{equation}
the absolute value of $\omega$ can be set equal to $1/\ell$ without loss of generality\footnote{This
corresponds to the choice made in case 8 of appendix \ref{stat-kill-bdry}.}.

\begin{figure}
\centerline{
\includegraphics[scale=0.7]{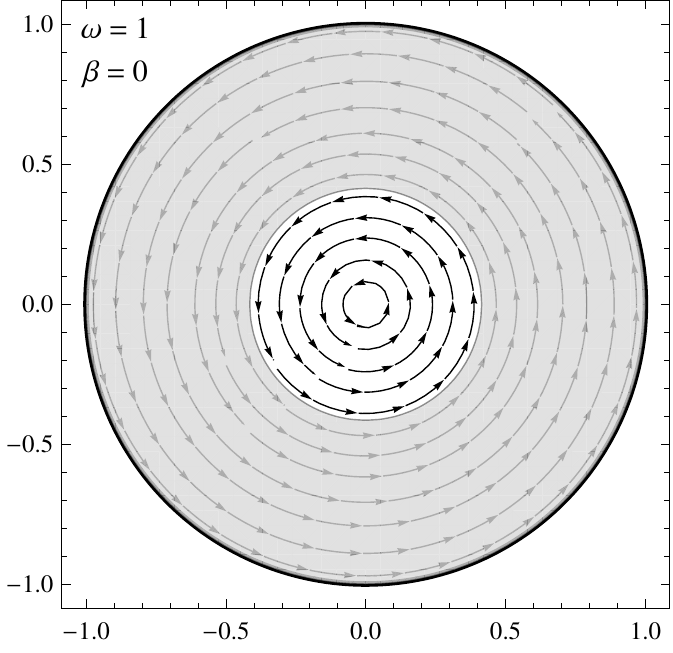}
\includegraphics[scale=0.7]{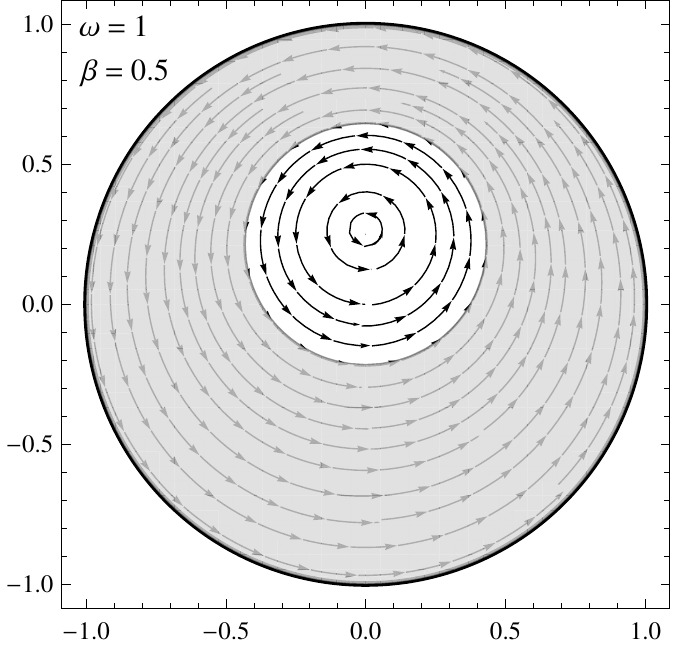}
\includegraphics[scale=0.7]{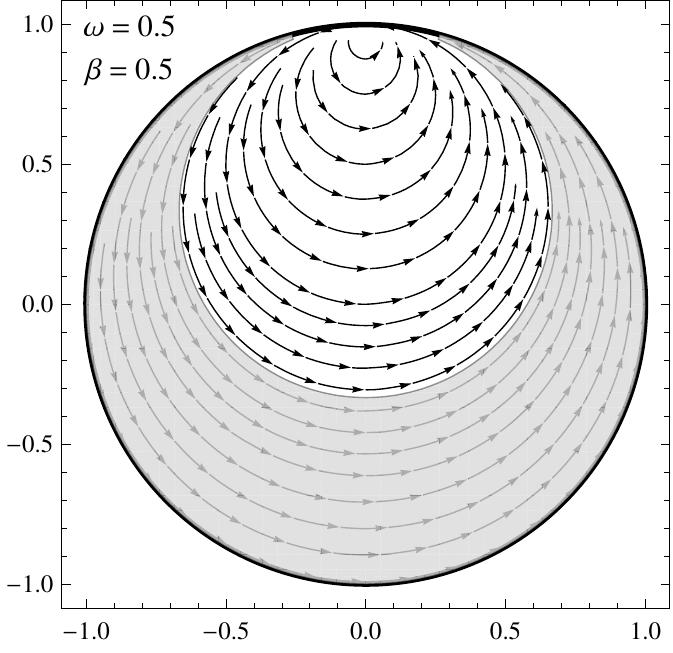}}
\centerline{
\includegraphics[scale=0.7]{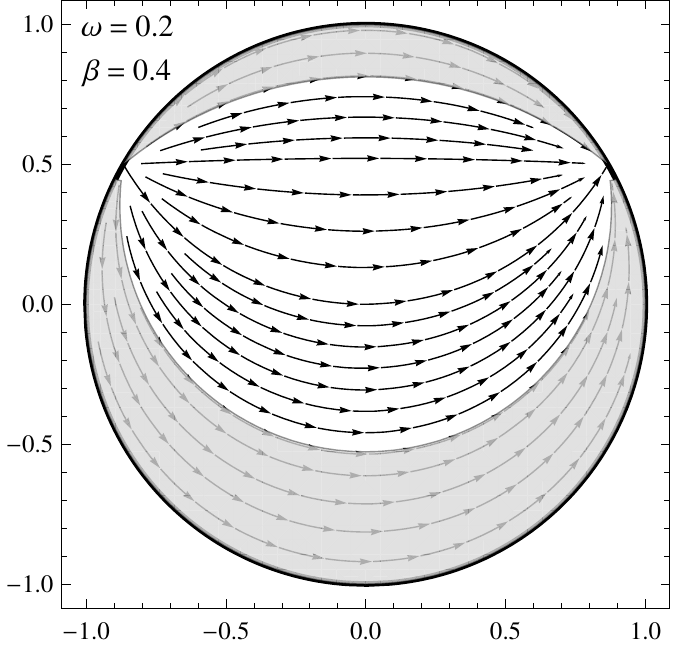}
\includegraphics[scale=0.7]{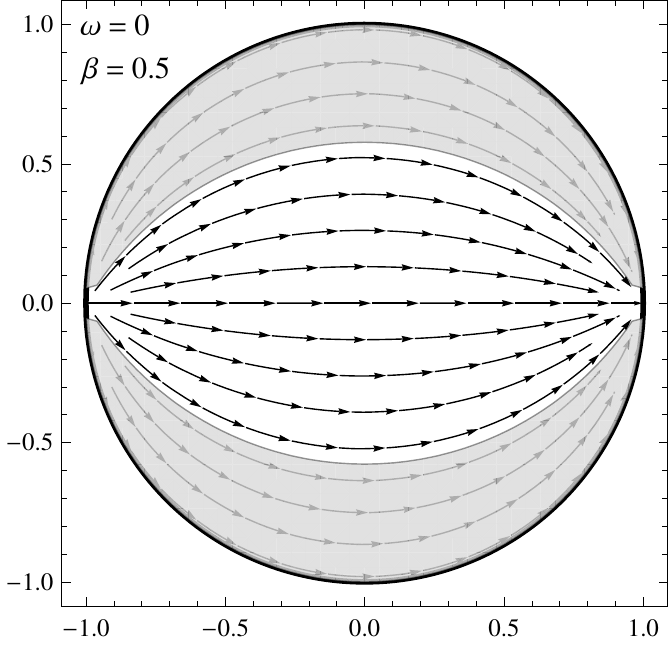}}
\caption{Integral curves (stream lines) of the vector field $v=\omega\xi^{(R)} + \beta\xi^{(B_1)}$ on the
Poincar\'e disk, for different values of $\omega$ and $\beta$. The white area denotes the physical
region, where the fluid velocity does not exceed the speed of light.}
\label{integral-curves}
\end{figure}

The integral curves of the fluid two-velocity $v=\omega\xi^{(R)} + \beta\xi^{(B_1)}$ are
visualized in figure \ref{integral-curves}. For $\omega^2>\beta^2$ the stream lines are closed and
the flow has one fixed point. For $\omega^2<\beta^2$ there are two fixed points lying on the
boundary of the Poincar\'e disk (which does not belong to the manifold itself).
If $\omega^2=\beta^2$, these fixed points coincide.
Of course, the cases $(\omega,\beta)=(1,0.5)$ and $(0.2,0.4)$ are isometric to $(1,0)$ and $(0,0.5)$
respectively.

In what follows, we shall analyze each of the three distinct cases separately.

\subsubsection{Rigid rotation}
\label{fluid rig rot hyp}

As was explained above, for $\omega^2>\beta^2$ one can take $\beta=0$ without loss of
generality. The 3-velocity of the fluid is then given by
\begin{equation}\label{rotating flux on the hyp}
u=\gamma(\partial_t+\omega\partial_\varphi)\,,
\end{equation}
where $\omega\in\mathbb{R}$ and $\gamma=(1-\omega^2\ell^2\sinh^2\theta)^{-1/2}$. Note that
the flow is well-defined only in the region $U=\{(\theta,\varphi)\,|\,|\omega|\ell\sinh\theta<1\}$.
At the boundary of $U$, the fluid rotates at the speed of light. Since $v=\omega\partial_\varphi$ is a
Killing field of $\text{H}^2$, this configuration is shearless and incompressible. The stress tensor
is given by
\begin{equation}\label{fluid on hyperboloid stress tensor}
T^{\mu\nu} = {\cal P}\begin{pmatrix}
3\gamma^2-1 & 0 & 3\gamma^2\omega\\ 
0 & \frac{1}{\ell^2} & 0\\ 
3\gamma^2\omega & 0 & \frac{3\gamma^2-2}{\ell^2\sinh^2\theta}
\end{pmatrix}\,,
\end{equation}
which is conserved once \eqref{P_0gamma^3} is satisfied. Moreover, the heat flux $q^\mu$
and diffusion current $q^\mu_{\text e}$ vanish by virtue of propositions \ref{prop-heat-flux} and
\ref{prop-diff-curr}.

Since the fluid velocity tends to the speed of light at the boundary of $U$, $\gamma$ diverges there
and the total energy and angular
momentum are infinite. Thus, unlike in the spherical case, we cannot define global thermodynamical
variables here, and have to consider instead only their densities. These are
\begin{enumerate}
\item the energy density $\varepsilon=T_{tt}={\cal P}_0\gamma^3(3\gamma^2-1)$,
\item the angular momentum density $l=-T_{t\varphi}=3{\cal P}_0\ell^2\omega\gamma^5\sinh^2\theta$,
\item the entropy density $\sigma=J_S^t=\tau^2\gamma^3(3h(\psi)-\psi h'(\psi))=\gamma s$,
\item the charge density $\varrho_{\text e}=J^t=\tau^2\gamma^3h'(\psi)=\gamma\rho_{\text e}$.
\end{enumerate}
We remark that these densities are evaluated in the frame $(t,\theta,\varphi)$, in which the fluid is moving, while the densities $\rho,s,\rho_{\text e}$ are measured in the local rest frame of the fluid.
Pointwise, $\varepsilon,l,\sigma$ and $\varrho_{\text e}$ are functions of the free parameters
$\omega,\tau,\psi$. Calculating their differentials one finds a local form of the first law,
\begin{equation}
\mathrm{d}\varepsilon = \tau\mathrm{d}\sigma + \omega\mathrm{d}l +
\tau\psi\mathrm{d}\varrho_{\text e}\,, \label{1st-law-local}
\end{equation}
which implies that the intensive variables conjugate to $\sigma,l$ and $\varrho_{\text e}$ are
respectively given by
\begin{equation}
\frac{\partial \varepsilon(\sigma,l,\varrho_{\text e})}{\partial \sigma} = \tau\,, \qquad
\frac{\partial \varepsilon(\sigma,l,\varrho_{\text e})}{\partial l} = \omega\,, \qquad
\frac{\partial \varepsilon(\sigma,l,\varrho_{\text e})}{\partial \varrho_{\text e}} = \tau\psi\,.
\end{equation}
The local grandcanonical potential
$\mathpzc{g}=\varepsilon-\tau\sigma-\omega l-\tau\psi\varrho_{\text e}$ reads
\begin{equation}
\mathpzc{g} = -\tau^3\gamma^3 h(\psi) = -\frac{\tau^3 h(\psi)}{(1-\omega^2\ell^2\sinh^2
\theta)^{3/2}}\,.
\end{equation}
\eqref{1st-law-local} is of course a consequence of local thermodynamical equilibrium.

\subsubsection{Purely translational flow} \label{sec:transl-flow}

Now we consider the case $\omega^2<\beta^2$, in which one can take $\omega=0$ without
loss of generality. This flow is visualized in the last figure of \ref{integral-curves}. In this case
it is convenient to use the coordinates
\begin{equation}
X = \sinh\theta\cos\varphi\,, \qquad Y = \sinh\theta\sin\varphi\,,
\end{equation}
in which the metric of the spacetime is given by
\begin{equation} \label{RxH2:XY}
g=-\mathrm{d}t^2+\frac{\ell^2}{1+X^2+Y^2}\left((1+Y^2)\mathrm{d}X^2+(1+X^2)\mathrm{d}Y^2-2XY\mathrm{d}X\mathrm{d}Y\right)\,,
\end{equation}
and the the fluid moves along the $X$ direction,
\begin{equation}
v=\beta\sqrt{1+X^2+Y^2}\partial_X\,.
\end{equation}
Since $v^2=\beta^2\ell^2(1+Y^2)$, the physical region $U$ is vertically narrowed by the condition
\begin{equation} 
Y^2<\frac{1}{\beta^2\ell^2}-1\,,
\end{equation}
which also shows that the flow exists only for $\beta^2<\ell^{-2}$. The 3-velocity reads
\begin{equation} 
u=\gamma(\partial_t+\beta\sqrt{1+X^2+Y^2}\partial_X)\,,
\end{equation}
where $\gamma=(1-\beta^2\ell^2(1+Y^2))^{-1/2}$.

Notice that the lower two figures of \ref{integral-curves} look very reminiscent of the black funnels
constructed in \cite{Fischetti:2012vt} to study heat transport in holographic CFT's. This raises the
question whether the bulk duals of the fluid flows in hyperbolic space considered here could be
used as toy models
for the gravity side of the construction in \cite{Fischetti:2012vt}. In this context, one should note
however that the black funnels of \cite{Fischetti:2012vt} contain a single connected bulk horizon
that extends to meet the conformal boundary. Thus the induced boundary metric has smooth horizons
as well. In our case instead, it turns out that the bulk horizon does not extend to meet the boundary,
although the
boundary metric itself may be considered to contain a horizon, since $\mathbb{R}\times\text{H}^2$
is conformal to the static patch of three-dimensional de~Sitter space \cite{Fischetti:2012vt}, which
has a cosmological horizon.

\subsubsection{Mixed flow: $\omega^2=\beta^2$} \label{sec:mixed-flow}

Finally, in the parabolic case $\omega^2=\beta^2$ one can choose $\omega=\beta$, as
was explained above. The Killing vector $v$ becomes then
\begin{displaymath}
v = \beta\left(iz + \frac12(1 - z^2)\right)\partial_z + \text{c.c.}
\end{displaymath}
It proves useful to introduce new coordinates $A,B$ defined by
\begin{displaymath}
A = \ln\frac{1 - z\bar z}{z\bar z + i(z - \bar z) + 1}\,, \qquad B = \frac{z + \bar z}{z\bar z +
i(z - \bar z) + 1}\,,
\end{displaymath}
such that $v=\beta\partial_B$ and
\begin{equation}
g=-\mathrm{d}t^2+\ell^2(\mathrm{d}A^2 + e^{-2A}\mathrm{d}B^2)\,.
\end{equation}
The 3-velocity becomes
\begin{equation}
u = \gamma(\partial_t + \beta\partial_B)\,,
\end{equation}
with the Lorentz factor $\gamma=(1-\beta^2\ell^2e^{-2A})^{-1/2}$. The physical region
$U$ is thus given by $1-\beta^2\ell^2e^{-2A}>0$.

\subsection{Fluid in rigid rotation on $\text{H}^2$ seen on the sphere or plane} \label{H2-to-plane}

The manifolds $\mathbb{R}\times\text{S}^2$ and $\mathbb{R}\times\text{H}^2$, with metrics
\eqref{metric sph spacetime} and \eqref{metric hyperb spacetime}, are conformally flat. This means
that each of them can be brought by a combined diffeomorphism plus Weyl rescaling
into a part of the other or into a part of three-dimensional Minkowski space $\mathbb{M}^3$.
One might thus ask how a fluid in one of these spaces appears when seen in the others after a conformal transformation. Since one may be interested in the description of hyperbolic AdS black holes
in terms of hydrodynamics on Minkowski space or on $\mathbb{R}\times\text{S}^2$,
we study as an example the rigidly rotating fluid on $\mathbb{R}\times\text{H}^2$ analyzed in subsection
\ref{fluid rig rot hyp} to see how it looks like on $\mathbb{M}^3$ or on the closed Einstein static
universe. We will see that this leads to interesting dynamical fluid configurations.

\begin{figure}
\centerline{
\includegraphics[scale=0.49]{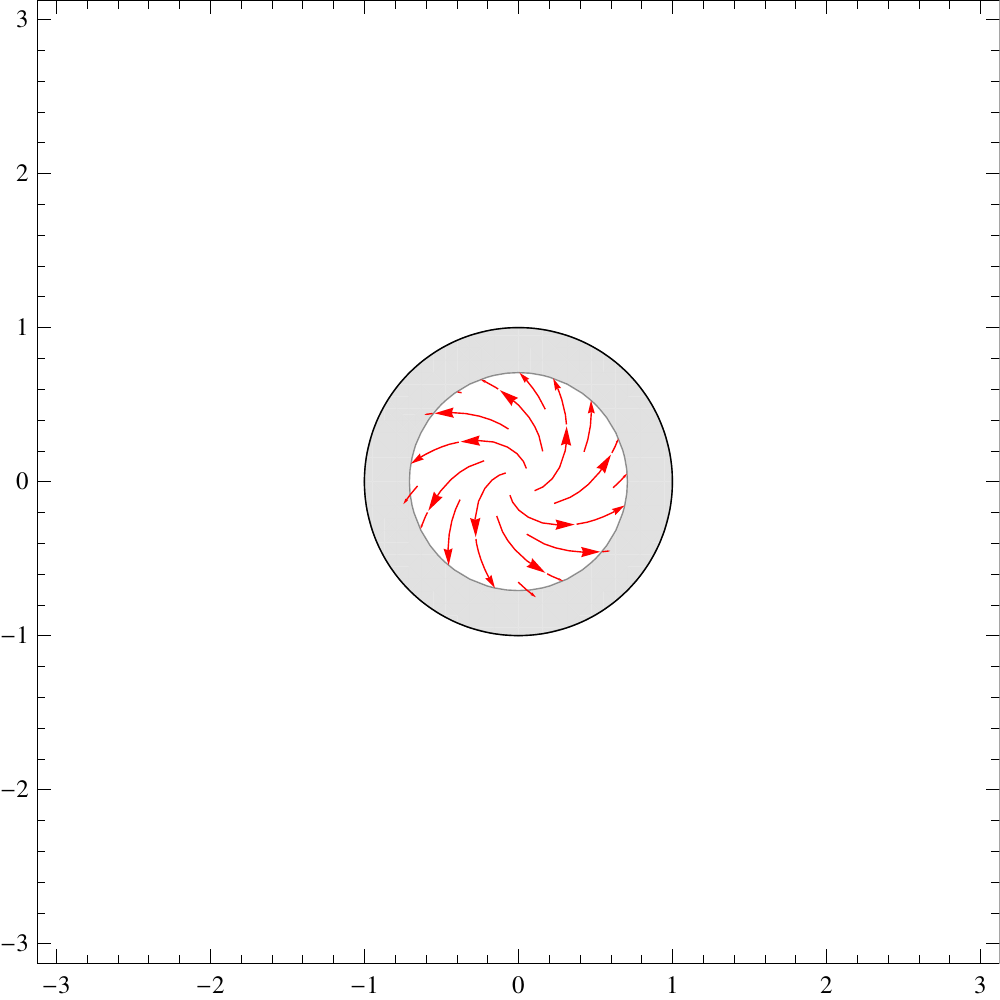}
\includegraphics[scale=0.49]{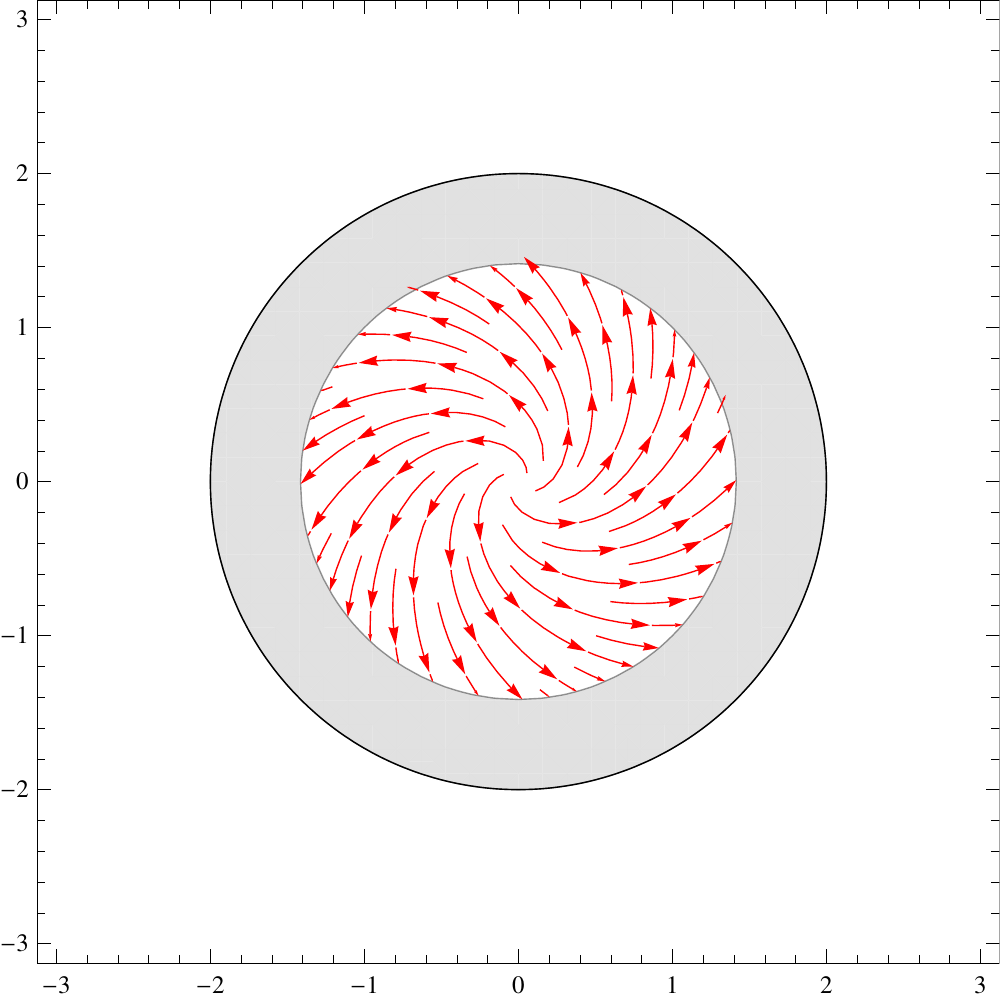}
\includegraphics[scale=0.49]{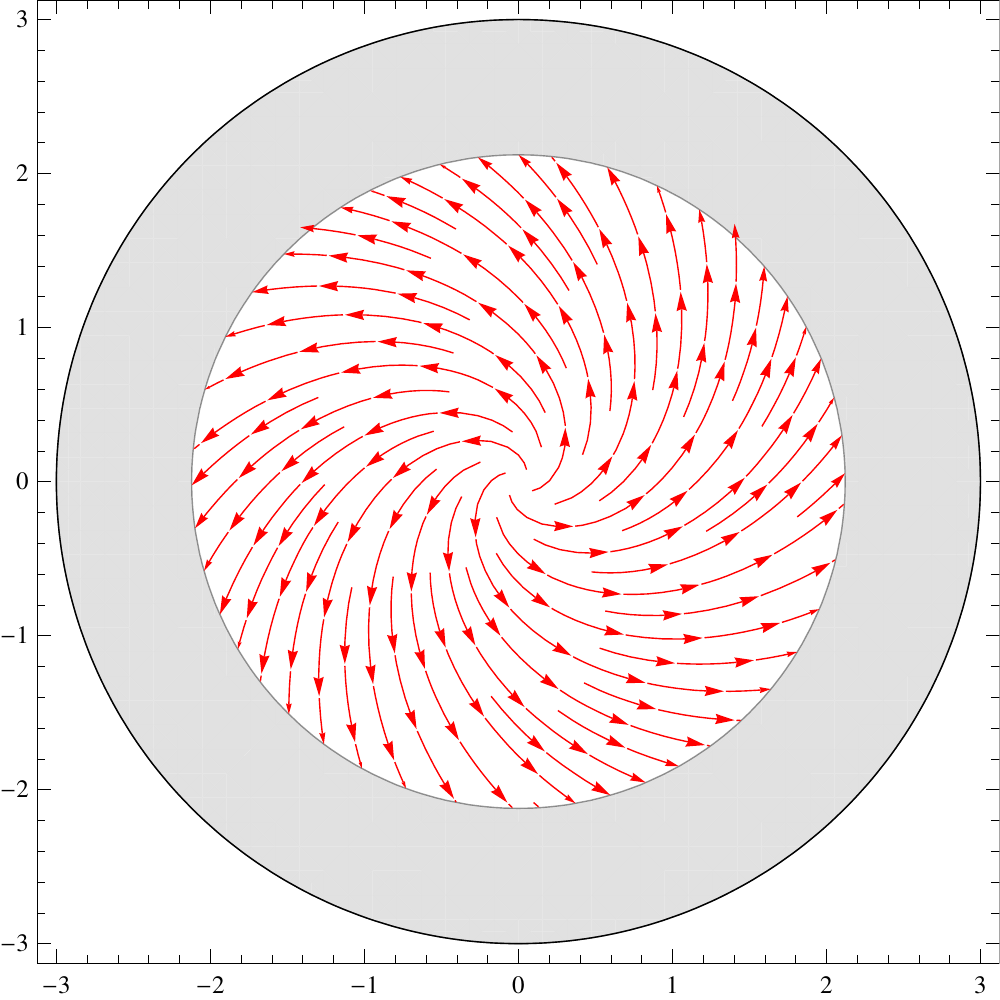}}
\caption{Fluid in rigid rotation on $\text{H}^2$ with $\omega=\ell=1$, seen on the plane in coordinates
$X,Y$, at times $T=1$, $T=2$ and $T=3$. The grey area is the region of spacetime where the flow is
not defined.}
\label{grafici R x H2 visto in R x E2}
\end{figure}

The coordinate transformation
\begin{equation} \label{H^2-to-Mink}
T = \ell e^{\frac t{\ell}}\cosh\theta\,, \qquad X = \ell e^{\frac t{\ell}}\sinh\theta\cos\varphi\,, \qquad
Y = \ell e^{\frac t{\ell}}\sinh\theta\sin\varphi\,,
\end{equation}
combined with a conformal rescaling $\tilde{g}=\Omega^2 g$, where
\begin{equation}
\Omega=e^\frac{t}{\ell}=\frac{\sqrt{T^2-X^2-Y^2}}{\ell}\,,
\end{equation}
brings \eqref{metric hyperb spacetime} to the flat metric
\begin{equation}
\tilde{g}=-\mathrm{d}T^2+\mathrm{d}X^2+\mathrm{d}Y^2\,.
\end{equation}
Now consider the rigidly rotating fluid in \ref{fluid rig rot hyp}, which has 3-velocity
\begin{equation}
u=\gamma(\partial_t+\omega\partial_\varphi)=
\frac{\gamma T}{\ell}\left(\partial_T+\frac{X-\omega\ell Y}{T}\partial_X+\frac{Y+\omega\ell X}{T}\partial_Y\right)\,,
\end{equation}
where
\begin{equation}
\gamma=(1-\omega^2\ell^2\sinh^2\theta)^{-\frac{1}{2}}=\sqrt{\frac{T^2-X^2-Y^2}{T^2-(1+\omega^2\ell^2)(X^2+Y^2)}}\,.
\end{equation}
Recall that the flow is defined only for $|\omega|\ell\sinh\theta<1$. In the
coordinates $(T,X,Y)$, this condition becomes $(1+\omega^2\ell^2)(X^2+Y^2)<T^2$.
Notice also that \eqref{H^2-to-Mink} maps $\mathbb{R}\times\text{H}^2$ to the inside of
the future light cone $X^2+Y^2<T^2$, $T>0$.
The conformal rescaling transforms $u$ into
\begin{equation}
\tilde{u} = \Omega^{-1}u = \frac{T}{\sqrt{T^2-(1+\omega^2\ell^2)(X^2+Y^2)}}
\left(\partial_T+\frac{X-\omega\ell Y}{T}\partial_X+\frac{Y+\omega\ell X}{T}\partial_Y\right)\,.
\end{equation}
This flow is plotted in coordinates $(T,X,Y)$ in figure \ref{grafici R x H2 visto in R x E2}.
We see that the rigidly rotating fluid in $\mathbb{R}\times\text{H}^2$ appears in Minkowski
space as an expanding vortex.

\begin{figure}
\centerline{
\includegraphics[scale=0.49]{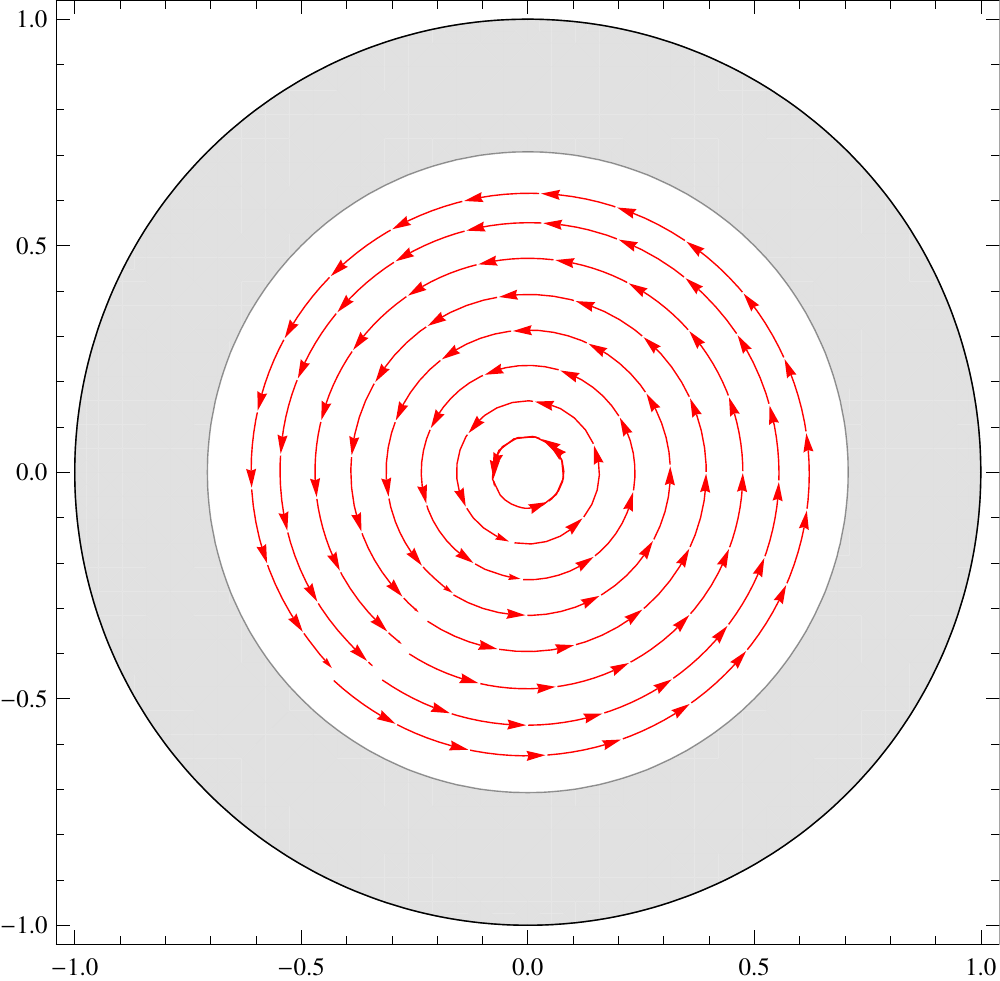}
\includegraphics[scale=0.49]{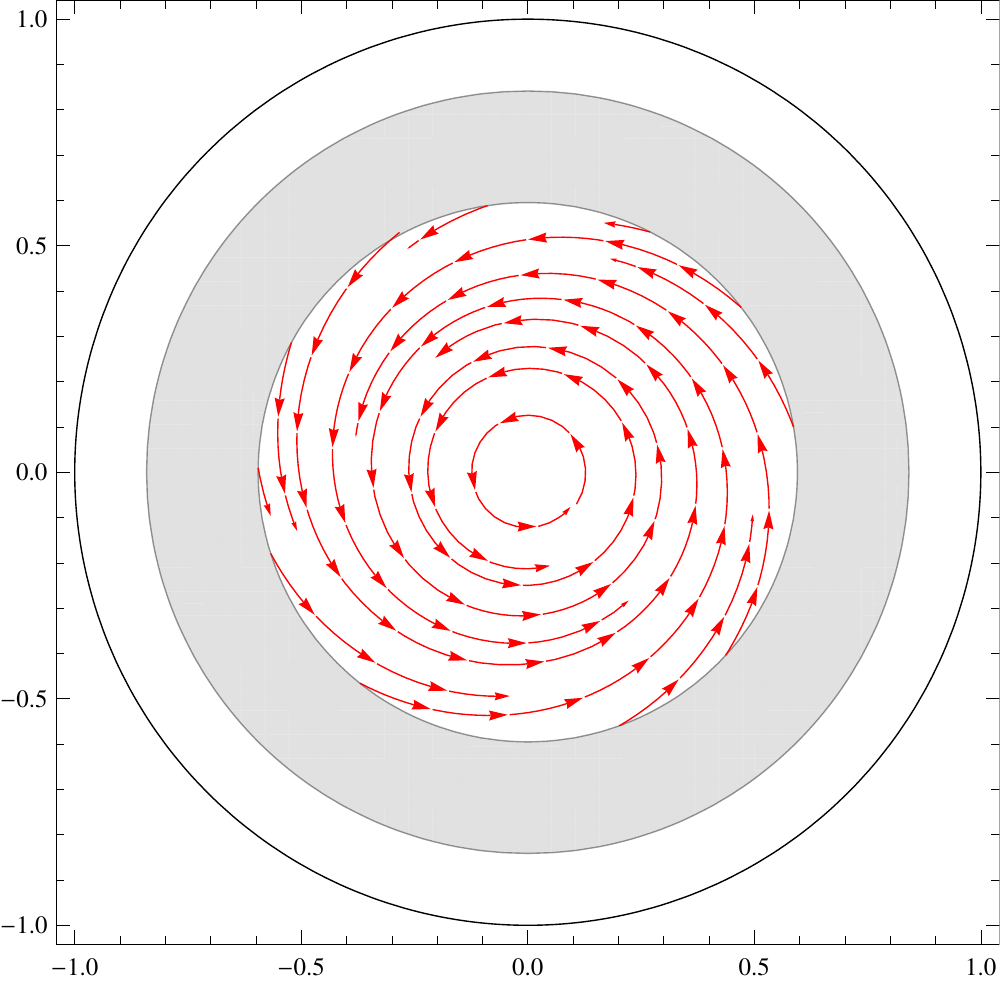}
\includegraphics[scale=0.49]{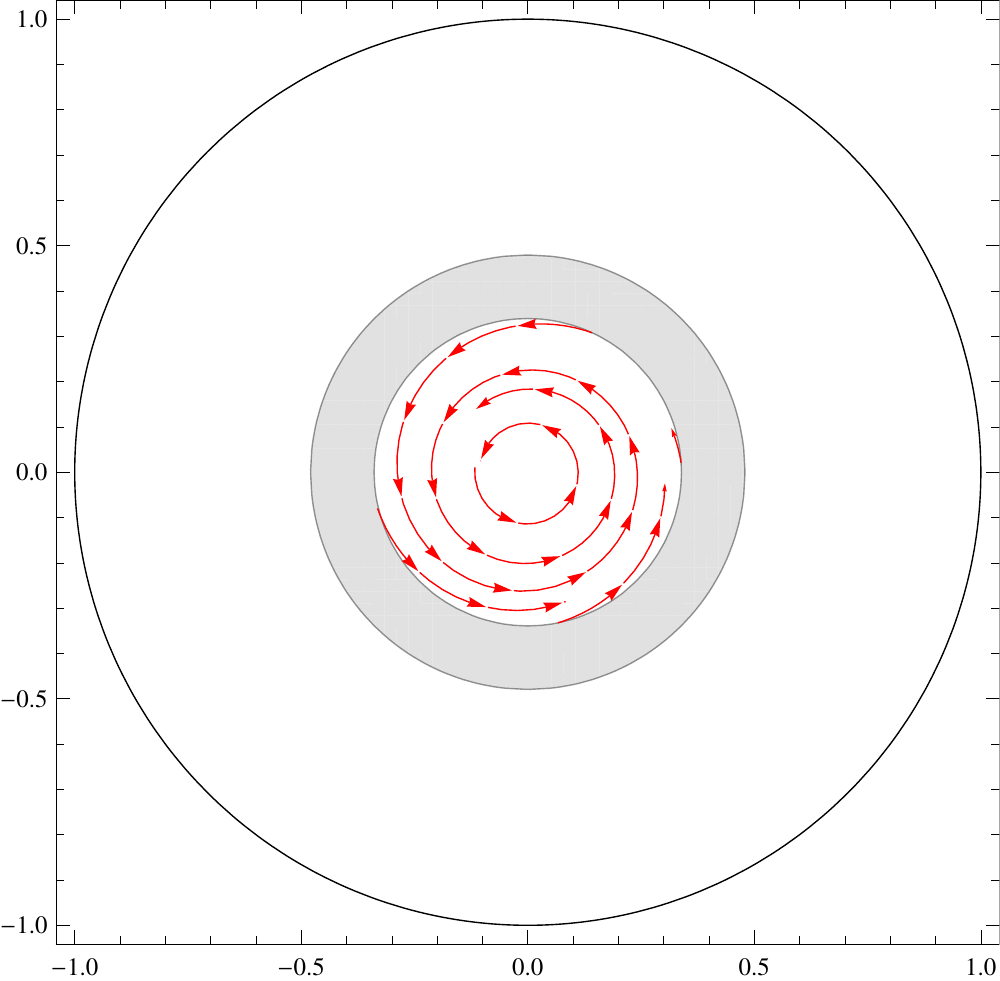}}
\caption{Fluid in rigid rotation on $\text{H}^2$ with $\omega=\ell=1$, seen on the 2-sphere (from
the north pole and projected on the equatorial plane), at times $\tau\simeq -\ell\frac{\pi}{2}$,
$\tau=-\ell$ and $\tau=-\frac{\ell}{2}$. The grey area is the region of spacetime where the flow is
not defined.}
\label{grafici R x H2 visto in R x S2}
\end{figure}

Let us now transform the same fluid configuration to the closed Einstein static universe
$\mathbb{R}\times\text{S}^2$. To this aim, introduce new coordinates
\begin{equation} \label{H^2-to-S^2}
\tau = -\ell \arctan\frac{\cosh\theta}{\sinh\frac{t}{\ell}}\,, \qquad
\Theta = \arctan\frac{\sinh\theta}{\cosh\frac{t}{\ell}}\,, \qquad
\Phi=\varphi\,,
\end{equation}
where $\tau\in(-\ell\frac{\pi}{2},0)$, $\Theta\in(0,\frac{\pi}{2})$ and $\Phi\in(0,2\pi)$. The inverse
of \eqref{H^2-to-S^2} is
\begin{equation}
t = \ell{\mathrm{arsinh}}\frac{\cos\frac{\tau}{\ell}}{\sqrt{\cos^2\Theta - \cos^2\frac{\tau}{\ell}}}\,,
\qquad \theta = {\mathrm{arsinh}}\frac{\sin\frac{\tau}{\ell}}{\sqrt{\cos^2\Theta -
\cos^2\frac{\tau}{\ell}}}\,,
\end{equation}
hence one has the additional restriction $\Theta<-\frac{\tau}{\ell}$. Subsequently, rescale
\eqref{metric hyperb spacetime} as $\tilde{g}=\Omega^2 g$, where
\begin{equation}
\Omega = \sqrt{\cos^2\Theta-\cos^2\frac{\tau}{\ell}}\,,
\end{equation}
to get
\begin{equation}
\tilde{g}=-\mathrm{d}\tau^2+\ell^2(\mathrm{d}\Theta^2+\sin^2\Theta\mathrm{d}\Phi^2)\,.
\end{equation}
Now the 3-velocity \eqref{rotating flux on the hyp} of the rigidly rotating fluid on $\text{H}^2$ is
mapped into
\begin{equation}
\tilde{u} = \Omega^{-1}u = \frac{-\sin\frac{\tau}{\ell}\cos\Theta}{\sqrt{\sin^2\frac{\tau}{\ell}-(1+\omega^2\ell^2)\sin^2\Theta}}
\left(\partial_\tau+\frac{\tan\Theta}{\ell\tan\frac{\tau}{\ell}}\partial_\Theta+\frac{\omega}{-\sin\frac{\tau}{\ell}\cos\Theta}\partial_\Phi\right)\,.
\end{equation}
In the coordinates $(\tau,\Theta,\Phi)$, the constraint $|\omega|\ell\sinh\theta<1$, limiting the region where the fluid is located, becomes
\begin{equation}
\sin\Theta<\frac{-\sin\frac{\tau}{\ell}}{\sqrt{1+\omega^2\ell^2}}\,.
\end{equation}
This flow is plotted in figure \ref{grafici R x H2 visto in R x S2} at different times $\tau$ projected on the equatorial plane of $\text{S}^2$ and viewed from the north pole.
Again, we encounter a dynamical fluid configuration that is a sort of contracting vortex on $\text{S}^2$.

Note that a similar technique was applied in 3+1 dimensions in \cite{Gubser:2010ui}. There,
it was shown (using the Weyl covariance of the stress tensor) that the dynamical solution of
\cite{Gubser:2010ze} (which represents a generalization of Bjorken flow \cite{Bjorken:1982qr}) can be
recast as a static flow in three-dimensional de~Sitter space times a line. The simplicity of the
de~Sitter form enabled the authors of \cite{Gubser:2010ui} to obtain several generalizations of it,
such as flows in other spacetime dimensions, second order viscous corrections, and linearized
perturbations.

\section{Dual AdS black holes} \label{dual-AdS-BH}

Now we want to identify the AdS black holes dual to the fluid configurations classified in
section \ref{stat flow ultrastatic st}. It turns out that these bulk spacetimes are all contained
in the Carter-Pleba\'nski family \cite{Carter:1968ks,Plebanski:1975}, whose metric is given by
\begin{equation} \label{Carter-Plebanski}
\mathrm{d}s^2 = \frac{p^2+q^2}{\mathsf{P}(p)}\mathrm{d}p^2+
\frac{\mathsf{P}(p)}{p^2+q^2}(\mathrm{d}\tau+q^2\mathrm{d}\sigma)^2+
\frac{p^2+q^2}{\mathsf{Q}(q)}\mathrm{d}q^2-
\frac{\mathsf{Q}(q)}{p^2+q^2}(\mathrm{d}\tau-p^2\mathrm{d}\sigma)^2\,,
\end{equation}
where
\begin{equation}
\mathsf{P}(p)=\alpha-\mathrm{g}^2+2lp-\epsilon p^2+\frac{p^4}{\ell^2}\,, \qquad
\mathsf{Q}(q)=\alpha+\mathrm{e}^2-2mq+\epsilon q^2+\frac{q^4}{\ell^2}\,. \label{PQ}
\end{equation}
This solves the Einstein-Maxwell equations with cosmological constant $\Lambda=-3\ell^{-2}$ and
electromagnetic field
\begin{equation}
A=-\frac{\mathrm{e}\,q}{p^2+q^2}(\mathrm{d}\tau-p^2\mathrm{d}\sigma)
-\frac{\mathrm{g}\,p}{p^2+q^2}(\mathrm{d}\tau+q^2\mathrm{d}\sigma)\,, \label{A-CP}
\end{equation}
whose field strength is
\begin{equation}
F=\frac{\mathrm{e}(p^2-q^2)+2\mathrm{g}\,pq}{(p^2+q^2)^2}\mathrm{d}q\wedge(\mathrm{d}\tau-p^2\mathrm{d}\sigma)
-\frac{\mathrm{g}(p^2-q^2)-2\mathrm{e}\,pq}{(p^2+q^2)^2}\mathrm{d}p\wedge(\mathrm{d}\tau+q^2\mathrm{d}\sigma)\,. \label{F-CP}
\end{equation}
\eqref{Carter-Plebanski} can be obtained by a scaling limit from the Pleba\'nski-Demia\'nski
spacetime \cite{Plebanski:1976gy}\footnote{This scaling limit eliminates the acceleration parameter.},
which is the most general known Petrov-type D solution to the
Einstein-Maxwell equations with cosmological constant. Other references studying algebraically
special spacetimes and their fluid duals include \cite{deFreitas:2014lia}, where the AdS/CFT interpretation
of the Robinson-Trautman (RT) solution to vacuum AdS gravity was investigated. This is slightly
different from our case, since the boundary metric of the RT geometry is in general
time-dependent \cite{deFreitas:2014lia}.

The metric $\hat g$ on the conformal boundary of \eqref{Carter-Plebanski} can be obtained by setting
$q=\text{const.}\to\infty$ and rescaling with $\ell^2/q^2$. This leads to
\begin{equation} \label{conf-bdry-CP}
{\hat g} = -\mathrm{d}\tau^2+
\frac{\ell^2}{\mathsf{P}(p)}\mathrm{d}p^2+
(\ell^2\mathsf{P}(p)-p^4)\mathrm{d}\sigma^2+2p^2\mathrm{d}\tau\mathrm{d}\sigma\,.
\end{equation}
Notice that for vanishing NUT-parameter $l$ this metric is conformally flat\footnote{The
nonvanishing components of the Cotton tensor $C_{\mu\nu\rho}$ of $\hat g$ are given by
\begin{displaymath}
C_{\tau p \sigma}=-C_{p\tau \sigma}=C_{\sigma \tau p}=-C_{\tau \sigma p}=\frac{2l}{\ell^2}\,, \qquad
C_{\sigma p \tau}=-C_{p \sigma \tau}=\frac{4l}{\ell^2}\,, \qquad
C_{p \sigma \sigma}=-C_{\sigma p \sigma}=\frac{6l p^2}{\ell^2}\,.
\end{displaymath}}. In what follows, we shall consider the case $l=0$ only\footnote{Holographic fluids
that are dual to geometries with NUT charge were considered
in \cite{Leigh:2011au,Caldarelli:2012cm,Mukhopadhyay:2013gja}.}.

Using standard holographic renormalization techniques \cite{Balasubramanian:1999re},
one can compute the holographic stress tensor associated to \eqref{Carter-Plebanski}, with the result
\begin{equation}
\hat{T}_{\mu\nu}=\frac{m}{8\pi\ell^2}(\gamma_{\mu\nu}+3u_\mu u_\nu)\,,
\end{equation}
where $u=\partial_\tau$. $\hat{T}$ describes thus a conformal fluid in equilibrium, at rest in the frame
$(\tau,p,\sigma)$.
The external electromagnetic field $\hat F$ and the $\text{U}(1)$ current $\hat J$ dual to
\eqref{A-CP} on the conformal boundary of \eqref{Carter-Plebanski} are found to be respectively
\begin{equation}
\hat{F} = \mathrm{g}\,\mathrm{d}p\wedge\mathrm{d}\sigma\,, \qquad
\hat{J} = \frac{\mathrm{e}}{4\pi\ell^2}\partial_\tau = \frac{\mathrm{e}}{4\pi\ell^2}u\,.
\end{equation}
The last equation shows that the fluid has a constant charge density $\mathrm{e}/(4\pi\ell^2)$.
Note also that the current $\hat{J}$ is conserved, ${\hat\nabla}_\mu\hat{J}^\mu=0$, where
$\hat\nabla$ denotes the Levi-Civita connection of $\hat g$. Moreover, since
$\hat{F}_{\mu\nu}\hat{J}^\nu=0$, the Lorentz force exerted by the field $\hat{F}$ on the
charged fluid vanishes, and thus $\hat T$ is conserved as well,
${\hat\nabla}_\mu\hat{T}^{\mu\nu}=0$.

Notice that the solution \eqref{Carter-Plebanski}, \eqref{F-CP} enjoys the scaling symmetry
\begin{eqnarray}
&&p\to \lambda p\,, \qquad q \to \lambda q\,, \qquad \tau \to \tau/\lambda\,, \qquad \sigma \to
\sigma/\lambda^3\,, \qquad \alpha\to \lambda^4\alpha\,, \nonumber \\
&&\mathrm{g}\to \lambda^2\mathrm{g}\,, \qquad
\mathrm{e} \to \lambda^2\mathrm{e}\,, \qquad
m\to \lambda^3 m\,, \qquad l \to \lambda^3 l\,, \qquad \epsilon \to
\lambda^2\epsilon\,, \label{scaling-symm}
\end{eqnarray}
that can be used to eliminate one unphysical parameter.

The line element \eqref{Carter-Plebanski} describes a black hole whose event horizon
$\cal H$ is located at the largest root of the polynomial $\mathsf{Q}(q)$. As we shall see below, the
horizon geometry depends crucially on the choice of parameters contained in the function
$\mathsf{P}(p)$, which determine the number of real roots of $\mathsf{P}$.
In what follows, we will discuss more in detail some subcases of the Carter-Pleba\'nski family,
which are dual to the fluid configurations classified in section \ref{stat flow ultrastatic st}.

\subsection{Spherical and hyperbolic Kerr-Newman-AdS$_4$ black holes}\label{Kerr-Newman-AdS4 BH}

If we set
\begin{displaymath}
\alpha = ka^2 + \mathrm{g}^2\,, \quad \epsilon = k + \frac{a^2}{\ell^2}\,, \quad \tau =
\frac{t - a\varphi}{\Xi}\,, \quad q = r\,, \quad p = a c_k(\theta)\,, \quad \sigma = -\frac{\varphi}{a\Xi}\,,
\end{displaymath}
where
\begin{displaymath}
k = \pm 1\,, \qquad \Xi = 1 - \frac{ka^2}{\ell^2}\,, \qquad c_k(\theta) = \frac{d s_k(\theta)}{d\theta}\,,
\qquad s_k(\theta)=\left\{\begin{matrix}
\sin\theta\,, \quad k=1\,, \\ 
\sinh\theta\,, \quad k=-1\,,
\end{matrix}\right.
\end{displaymath}
the metric \eqref{Carter-Plebanski} becomes
\begin{equation}
\mathrm{d}s^2 = -\frac{\Delta_r}{\Xi^2\rho^2}\left(\mathrm{d}t-ka\,s_k^2(\theta)\mathrm{d}
\varphi\right )^2 + \rho^2\left(\frac{\mathrm{d}r^2}{\Delta_r}+
\frac{\mathrm{d}\theta^2}{\Delta_\theta}\right )+
\frac{\Delta_\theta}{\Xi^2\rho^2}\left(a\mathrm{d}t-(r^2+a^2)\mathrm{d}\varphi\right )^2
s_k^2(\theta)\,, \label{sph-hyp KNAdS4}
\end{equation}
with
\begin{displaymath}
\rho^2 = r^2+a^2 c_k^2(\theta)\,, \qquad \Delta_r = (r^2+a^2)\left(k+\frac{r^2}{\ell^2}\right) - 2mr
+ \mathrm{e}^2 + \mathrm{g}^2\,, \qquad \Delta_\theta = 1 - \frac{ka^2}{\ell^2}c_k^2(\theta)\,.
\end{displaymath}
For $k=1$ this is the Kerr-Newman-AdS$_4$ black hole, while for $k=-1$ one has the rotating
hyperbolic solution constructed in \cite{Klemm:1997ea}. Note also that in the spherical case
$(k=1)$ the rotation parameter $a$ is bounded by $a^2<\ell^2$ in order for $\Delta_{\theta}$ to be
positive, while it can take any value if $k=-1$.

\begin{figure}
\centerline{
\includegraphics[scale=0.45]{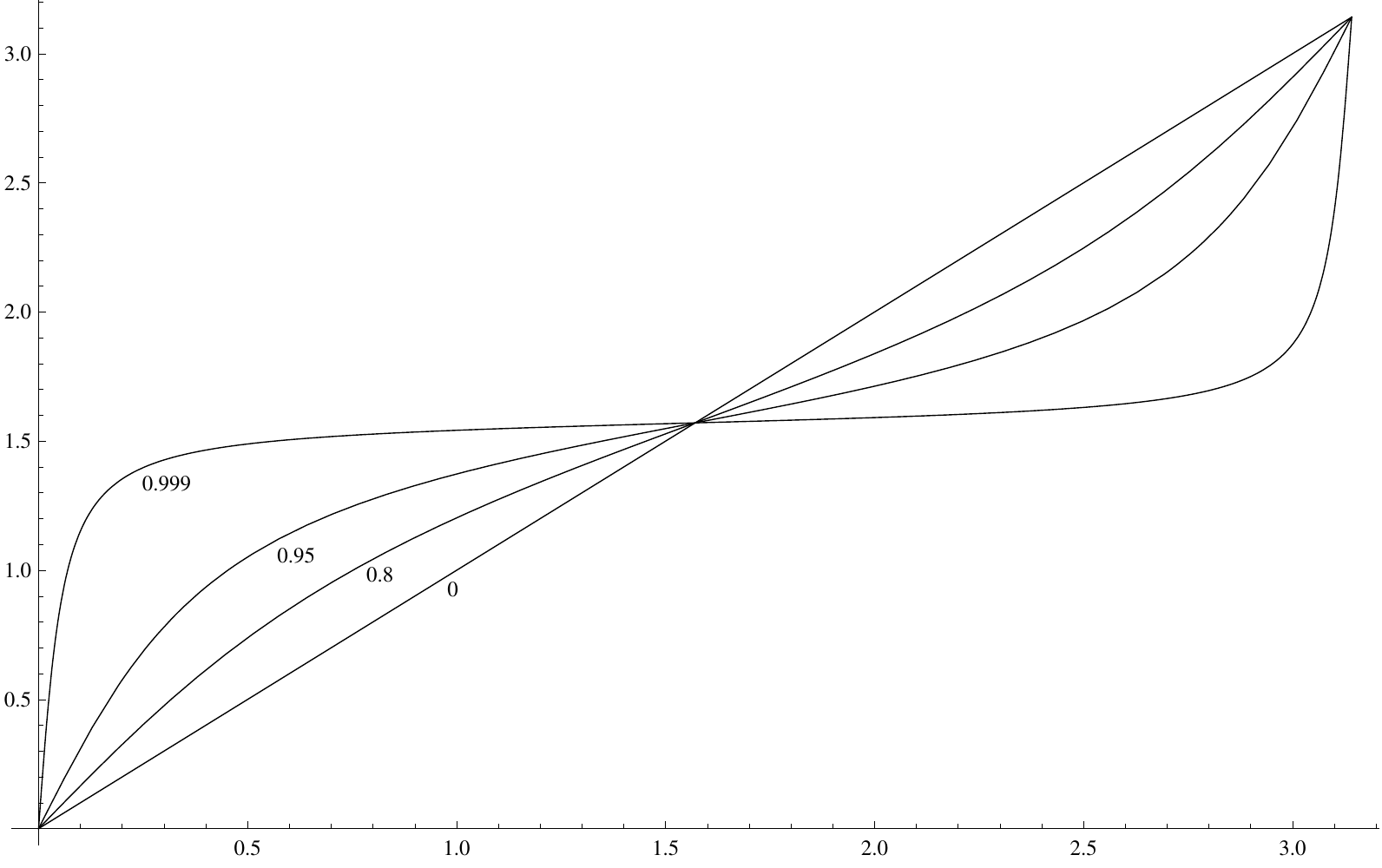}
\includegraphics[scale=0.45]{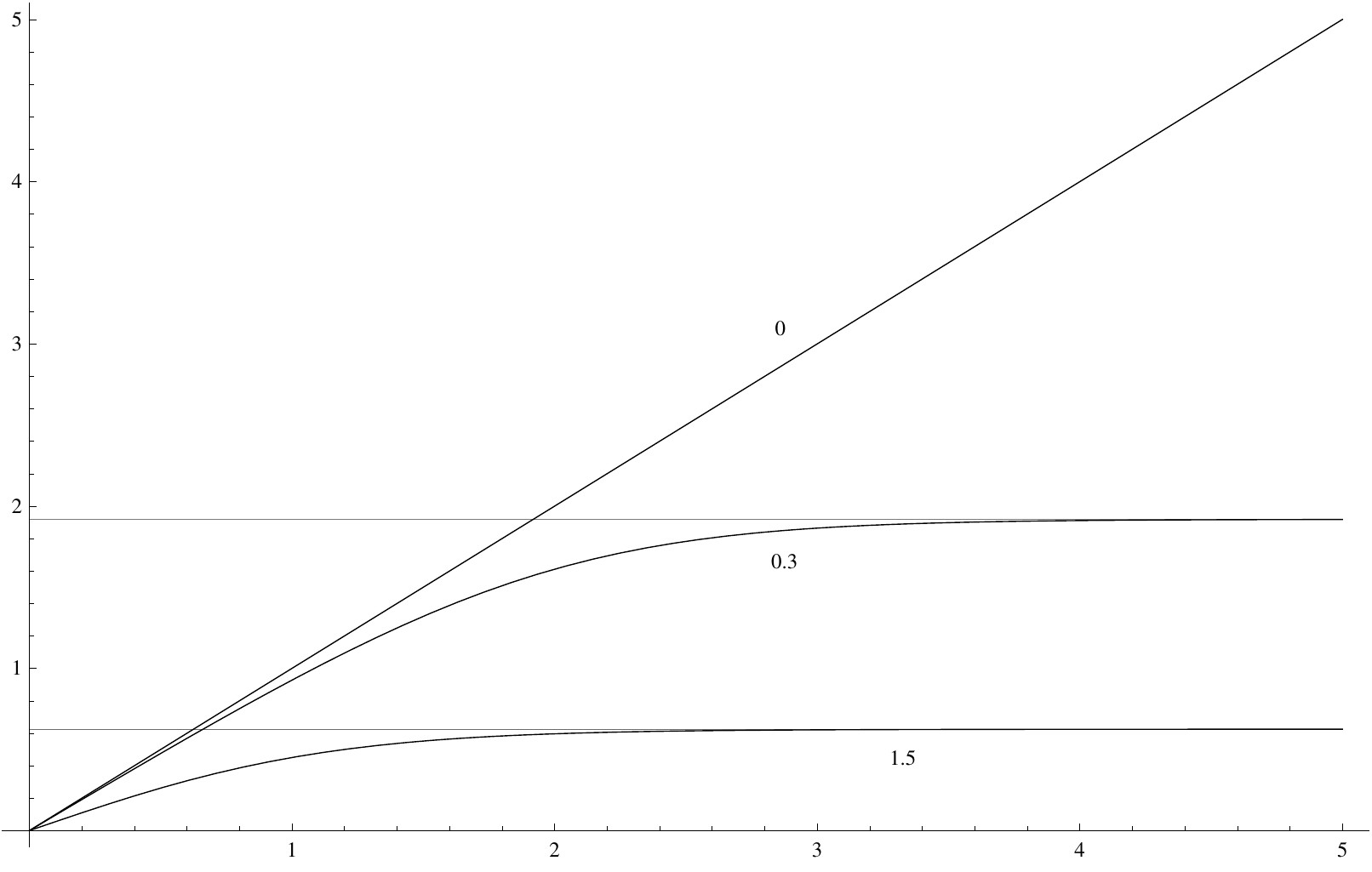}}
\caption{Graphs of $\Theta(\theta)$ for $k=1$ (left) and $k=-1$ (right), for different values of $a/\ell$.}
\label{grafici Theta(theta)}
\end{figure}

The metric on the conformal boundary of \eqref{sph-hyp KNAdS4} reads
\begin{equation}\label{metric conf bound s-h rotating bh}
{\hat g} = -\frac{\mathrm{d}t^2}{\Xi^2}+\frac{\ell^2\mathrm{d}\theta^2}{\Delta_{\theta}}+
\frac{\ell^2}{\Xi}s_k^2(\theta)\mathrm{d}\varphi^2+2\frac{ak}{\Xi^2}s_k^2(\theta)\mathrm{d}t\mathrm{d}\varphi\,.
\end{equation}
Since this is conformally flat there exist coordinates in which, after a conformal rescaling, it takes the
ultrastatic spherical or hyperbolic form (like in eqns.~\eqref{metric sph spacetime} and
\eqref{metric hyperb spacetime}). These are given by
\begin{equation}\label{coord change conf bound}
\tau = \frac{t}{\Xi}\,, \qquad c_k(\Theta) = c_k(\theta)\sqrt{\frac{\Xi}{\Delta_\theta}}\,, \qquad
\Phi = \varphi+\frac{kat}{\ell^2\Xi}\,.
\end{equation}
Notice that $\Theta$ ranges in $(0,\pi)$ when $k=1$ and in $(0,\text{arsinh}(\ell/|a|))$
when $k=-1$, cf.~fig.~\ref{grafici Theta(theta)}, where $\Theta(\theta)$ is plotted for different values
of $a/\ell$.
In the new coordinates, the boundary metric \eqref{metric conf bound s-h rotating bh} takes the form
\begin{equation}
{\hat g} = \frac{\Delta_\theta}{\Xi}\left(-\mathrm{d}\tau^2+\ell^2(\mathrm{d}\Theta^2+s_k^2(\Theta)\mathrm{d}\Phi^2)\right)\,,
\end{equation}
such that, after a Weyl rescaling
\begin{equation}
{\hat g}\rightarrow\tilde{g}=\Omega^2{\hat g}\,, \qquad
\Omega^2=\Xi/\Delta_{\theta}\,, \label{Weyl-KNAdS}
\end{equation}
one obtains the desired metric
\begin{equation}\label{rescaled metr on bound}
\tilde{g} = -\mathrm{d}\tau^2+\ell^2(\mathrm{d}\Theta^2+s_k^2(\Theta)\mathrm{d}\Phi^2)\,.
\end{equation}
Thus the boundary of the Kerr-AdS$_4$ black hole is conformal to $\mathbb{R}\times\text{S}^2$
for $k=1$, and to the part of $\mathbb{R}\times\text{H}^2$ with $\sinh\Theta<\ell/|a|$ for $k=-1$.
If we identify $|\omega|=|a|/\ell^2$, this is exactly the part of $\mathbb{R}\times\text{H}^2$ on which
a fluid in rigid rotation with angular velocity $\omega$ does not exceed the speed of light. We will have
to say more on this below.

The holographic stress tensor associated to \eqref{sph-hyp KNAdS4} can be written in the form
\begin{equation}\label{black hole stress tensor}
\hat{T}_{\mu\nu}=\frac{m}{8\pi\ell^2}\left({\hat g}_{\mu\nu}+3u_\mu u_\nu \right )\,,
\end{equation}
where $u=\Xi\partial_t$. This is the stress tensor of a conformal fluid at rest in the coordinate frame
$(t,\theta,\varphi)$, with pressure ${\cal P}=m/(8\pi\ell^2)$. After the diffeomorphism
\eqref{coord change conf bound} and the subsequent Weyl rescaling \eqref{Weyl-KNAdS} (recall that
$\hat T$ transforms as $\tilde T^{\mu\nu}=\Omega^{-d-2}\hat T^{\mu\nu}$) one obtains
\begin{equation}
\tilde{T}^{\mu\nu}=\frac{m\gamma^3}{8\pi\ell^2}
\begin{pmatrix}
3\gamma^2-1 & 0 & \frac{3ka \gamma^2}{\ell^2}\\ 
0 & \frac{1}{\ell^2} & 0\\ 
\frac{3ka \gamma^2}{\ell^2} & 0 & \frac{3\gamma^2-2}{\ell^2 s_k^2(\Theta)}
\end{pmatrix}\,, \label{tilde-T-KNAdS}
\end{equation}
with $\gamma:=(1-a^2s_k^2(\Theta)/\ell^2)^{-1/2}$. This can also be rewritten as
$\tilde{T}^{\mu\nu}=\tilde{\cal P}\left(\tilde{g}^{\mu\nu}+3\tilde{u}^\mu \tilde{u}^\nu\right )$,
where
\begin{equation}
\tilde{\cal P}=\Omega^{-3}{\cal P} = \frac{m\gamma^3}{8\pi\ell^2}\,, \qquad \tilde{u} =\Omega^{-1}u
= \gamma(\partial_\tau+\frac{ka}{\ell^2}\partial_\Phi)\,.
\end{equation}
For $k=1$, $\tilde{T}$ is exactly the stress tensor \eqref{fluid on sphere stress tensor} of the stationary conformal fluid on the 2-sphere, if we identify
\begin{equation}\label{spher corr rules}
{\cal P}_0 = \frac{m}{8\pi\ell^2}\,, \qquad \omega = \frac{a}{\ell^2}\,.
\end{equation}
On the other hand, for $k=-1$, \eqref{tilde-T-KNAdS} coincides with the stress tensor
\eqref{fluid on hyperboloid stress tensor} of the rigidly rotating conformal fluid on the hyperbolic plane,
after the identifications
\begin{equation}\label{hyp corr rules}
{\cal P}_0= \frac{m}{8\pi\ell^2}\,, \qquad \omega = -\frac{a}{\ell^2}\,.
\end{equation}
The KNAdS black hole is thus dual to a fluid in rigid rotation on $\text{S}^2$ for $k=1$ and on
$\text{H}^2$ for $k=-1$. In the spherical case, this is of course
well-known \cite{Bhattacharyya:2007vs,Caldarelli:2008ze}.
The result for hyperbolic
black holes is new, and it is remarkable how the conformal transformation \eqref{coord change conf bound},
\eqref{Weyl-KNAdS} maps the boundary geometry of the rotating hyperbolic black hole precisely
to the region of $\mathbb{R}\times\text{H}^2$ on which a fluid in rigid rotation does not exceed the
speed of light.

The electromagnetic field and electric current on the boundary are given respectivey by
\begin{equation}
\hat{F} = \frac{k\mathrm{g}s_k(\theta)}{\Xi}\mathrm{d}\theta\wedge\mathrm{d}\varphi\,, \qquad
\hat{J} = \frac{\mathrm{e}\Xi}{4\pi\ell^2}\partial_t = \frac{\mathrm{e}}{4\pi\ell^2}u\,.
\end{equation}
After the coordinate change \eqref{coord change conf bound} and Weyl rescaling \eqref{Weyl-KNAdS},
they become\footnote{One has $\tilde F_{\mu\nu}=\hat F_{\mu\nu}$ and
$\tilde J^\mu=\Omega^{-d}\hat J^\mu$. In this way, the MHD equations \eqref{eq:MHD} are
conformally invariant.}
\begin{equation}
\tilde{F} = k\mathrm{g}\gamma^3 s_k(\Theta)\,\mathrm{d}\Theta\wedge(\mathrm{d}\Phi-
\frac{ka}{\ell^2}\mathrm{d}\tau)\,, \qquad
\tilde{J} = \frac{\mathrm{e}\gamma^3}{4\pi\ell^2}(\partial_\tau+\frac{ka}{\ell^2}\partial_\Phi)
= \frac{\mathrm{e}\gamma^2}{4\pi\ell^2}\tilde{u}\,,
\end{equation}
and thus $\tilde J$ coincides with the hydrodynamical expression if the charge density of the fluid is
\begin{equation}
\rho_{\text e} = \frac{\mathrm{e}\gamma^2}{4\pi\ell^2}\,. \label{rho_e}
\end{equation}
Note that in the coordinate system $(\tau,\Theta,\Phi)$ there is also an electric field. Moreover, one
has $\tilde{F}^\nu{}_\mu\tilde{J}^\mu=0$, so there is no net Lorentz force acting on the
charged fluid due to an exact cancellation of electric and magnetic forces\footnote{In the case of
the spherical KNAdS black hole this fact was first noticed in \cite{Caldarelli:2008ze}.}.
In the orthonormal frame
\begin{displaymath}
e^0 = \mathrm{d}\tau\,, \qquad e^1 = \ell\mathrm{d}\Theta\,, \qquad e^2 = \ell s_k(\Theta)
\mathrm{d}\Phi\,,
\end{displaymath}
the electric field in $1$-direction and the spatial current in $2$-direction are
\begin{equation}
E^1 = \frac{\mathrm{g}\gamma^3 s_k(\Theta)a}{\ell^3}\,, \qquad \tilde{J}^{\,2} =
\frac{\mathrm{e}\gamma^3 s_k(\Theta)ka}{4\pi\ell^3} = \sigma^{21}E^1\,,
\end{equation}
with the Hall conductivity
\begin{equation}
\sigma^{21} = \frac{\mathrm{e}k}{4\pi\mathrm{g}}\,.
\end{equation}

In the spherical case $k=1$, it was furthermore shown in \cite{Bhattacharyya:2007vs} that, in the large
black hole limit where fluid dynamics provides an accurate description of the dual conformal field
theory, the black hole electric charge, entropy, mass and
angular momentum coincide precisely with the conserved charges \eqref{Q-S}, \eqref{E}
and \eqref{L} computed in fluid mechanics, if we identify the Hawking temperature $T$ with the
global fluid temperature in \eqref{T-Omega-phi}\footnote{The remaining fluid parameters
$\omega$ and $\psi$ are fixed by \eqref{spher corr rules} and \eqref{rho_e} (combined with
$\rho_{\text e}=\tau^2\gamma^2h'(\psi)$, cf.~\eqref{P_0-rho_e-s}) respectively. The function
$h(\psi)$ determining the hydrodynamic grandcanonical potential is that of the unrotating
black hole, given by eqns.~(3.19) and (3.20) of \cite{Caldarelli:2008ze}.}.

On the other hand, for $k=-1$, we already saw in the previous section that the conserved charges
are ill-defined in fluid mechanics. The same problem is encountered on the gravity side:
If one tries to compute for instance the entropy of the solution \eqref{sph-hyp KNAdS4} with $k=-1$,
one has to integrate over the noncompact variable $\theta$, which makes the result divergent.
A possible way out could be to consider only excitations above some `ground state', which may
have finite energy, but we shall not attempt to do this here. In spite of these difficulties, we saw in
section \ref{fluid rig rot hyp} that a local form of the first law of black hole mechanics holds.

\subsection{Boosting AdS$_4$ black holes}\label{boosting charged bh}

In the previous subsection it was shown that the spherical and hyperbolic KNAdS$_4$ black holes
are holographically dual to conformal fluids in rigid rotation on $\mathbb{R}\times\text{S}^2$
and $\mathbb{R}\times\text{H}^2$ respectively. While a rigid rotation is (up to isometries) the only
possible equilibrium configuration for a stationary conformal fluid on a sphere, the same is not true for
hyperbolic space: We saw in \ref{sec:fluid-H2} that on the hyperbolic plane one can
also have purely translational (`boosting') or mixed flows, which are not isometric to rotations. In this
section we describe a family of black holes, obtained by analytically continuing the hyperbolic
KNAdS$_4$ metric, whose dual fluid is translating on the hyperbolic plane. We will call these
solutions `boosting black holes'.

Consider the KNAdS$_4$ metric \eqref{sph-hyp KNAdS4} with $k=-1$, and analytically continue
\begin{equation}\label{anal cont KNAdS metric}
a\rightarrow ib\,, \qquad \theta\rightarrow\theta-\frac{i\pi}{2}\,, \qquad \varphi\rightarrow
i\varphi\,.
\end{equation}
This leads to
\begin{equation}
\mathrm{d}s^2 = -\frac{\Delta_r}{\Xi^2\rho^2}\left(\mathrm{d}t+b\cosh^2\theta\mathrm{d}\varphi
\right )^2 + \rho^2\left(\frac{\mathrm{d}r^2}{\Delta_r} + \frac{\mathrm{d}\theta^2}{\Delta_\theta}
\right ) + \frac{\Delta_\theta\cosh^2\theta}{\Xi^2\rho^2}\left(b\mathrm{d}t-(r^2-b^2)
\mathrm{d}\varphi\right )^2\,,\label{boostingAdS4}
\end{equation}
where now
\begin{displaymath}
\rho^2 = r^2+b^2\sinh^2\theta\,, \quad
\Delta_r = (r^2-b^2)\left(-1+\frac{r^2}{\ell^2}\right) - 2m r + \mathrm{e}^2 + \mathrm{g}^2\,,
\quad \Delta_{\theta} = 1+\frac{b^2}{\ell^2}\sinh^2\theta\,,
\end{displaymath}
and $\Xi=1-b^2/\ell^2$. Alternatively, \eqref{boostingAdS4} can be obtained directly from the
Carter-Pleba\'nski solution \eqref{Carter-Plebanski} by setting
\begin{displaymath}
\alpha = b^2 + \mathrm{g}^2\,, \quad l = 0\,, \quad \epsilon = -1-\frac{b^2}{\ell^2}\,, \quad
\tau = \frac{t+b\varphi}{\Xi}\,, \quad q = r\,, \quad p = b\sinh\theta\,, \quad \sigma =
-\frac{\varphi}{b\Xi}\,.
\end{displaymath}
The electromagnetic 1-form potential \eqref{A-CP} becomes then
\begin{equation}
A = -\frac{\mathrm{e}r}{\Xi\rho^2}\left(\mathrm{d}t+b\cosh^2\theta\mathrm{d}\varphi\right)
-\frac{\mathrm{g}\sinh\theta}{\Xi\rho^2}\left(b\mathrm{d}t-(r^2-b^2)\mathrm{d}\varphi\right)\,.
\label{A-boostingAdS4}
\end{equation}
Notice that now $\theta,\varphi$ are not polar coordinates on $\text{H}^2$
(in that case it would not be possible to extend the 1-form $\cosh\theta\,\mathrm{d}\varphi$ to
$\theta=0$), but they are rather Cartesian-type coordinates on a plane, possibly compactified to a
cylinder by periodic identifications of $\varphi$\footnote{In the latter case the dual fluid lives
on a quotient space of $\text{H}^2$.}.

The metric on the conformal boundary of \eqref{boostingAdS4} is given by
\begin{equation}
{\hat g} = -\frac{\mathrm{d}t^2}{\Xi^2}+\frac{\ell^2}{\Delta_\theta}\mathrm{d}\theta^2+\frac{\ell^2}
{\Xi}\cosh^2\theta\mathrm{d}\varphi^2-2\frac{b}{\Xi^2}\cosh^2\theta\mathrm{d}t\mathrm{d}\varphi\,,
\end{equation}
from which we see that $\partial_{\varphi}$ is spacelike only for $b^2<\ell^2$. Now introduce the
ultrastatic coordinates
\begin{equation} \label{coord change conf bound mod}
T = \frac{t}{\Xi}\,, \qquad X = \frac{\cosh\theta}{\sqrt{\Delta_\theta}}\sinh\left(\varphi-\frac{bt}
{\ell^2\Xi}\right)\,, \qquad Y = \sqrt{\frac{\Xi}{\Delta_\theta}}\sinh\theta\,,
\end{equation}
where $T,X\in\mathbb{R}$ and $Y$ is bounded by $Y^2<\ell^2/b^2-1$, and perform a Weyl
rescaling $\tilde{g}=\Omega^2{\hat g}$ with
\begin{equation}
\Omega = \sqrt{\frac{\Xi}{\Delta_\theta}} = \sqrt{1-\frac{b^2}{\ell^2}(1+Y^2)}\,. \label{Weyl-boost}
\end{equation}
This yields
\begin{equation}\label{Gamma mod}
\tilde{g} = -\mathrm{d}T^2+\frac{\ell^2}{1+X^2+Y^2}\left((1+Y^2)\mathrm{d}X^2+(1+X^2)
\mathrm{d}Y^2-2XY\mathrm{d}X\mathrm{d}Y\right)\,,
\end{equation}
which is the slice $Y^2<\ell^2/b^2-1$ of the spacetime $\mathbb{R}\times\text{H}^2$,
cf.~\eqref{RxH2:XY}.

In the frame $(t,\theta,\varphi)$ the holographic stress tensor on the conformal boundary is found to be
\begin{equation}
\hat{T}_{\mu\nu}=\frac{m}{8\pi\ell^2}\left(\gamma_{\mu\nu}+3u_\mu u_\nu \right )\,,
\end{equation}
with $u=\Xi\partial_t$. After the diffeomorphism \eqref{coord change conf bound mod} and the Weyl
rescaling \eqref{Weyl-boost}, the stress tensor becomes
\begin{equation}
\tilde{T}_{\mu\nu} = \tilde{\cal P}\left(\tilde{g}_{\mu\nu}+3\tilde{u}_\mu\tilde{u}_\nu\right )\,,
\end{equation}
where
\begin{displaymath}
\tilde{\cal P} = \frac{m\gamma^3}{8\pi\ell^2}\,, \qquad
\tilde{u} = \gamma\left(\partial_T-\frac{b}{\ell^2}\sqrt{1+X^2+Y^2}\partial_X\right)\,, \qquad
\gamma = \Omega^{-1} = \left(1-\frac{b^2}{\ell^2}(1+Y^2)\right)^{-1/2}\,.
\end{displaymath}
This is exactly the energy-momentum tensor and 3-velocity of a conformal fluid translating on the
hyperbolic plane studied in section \ref{sec:transl-flow}, after the identifications
\begin{equation}
{\cal P}_0 = \frac{m}{8\pi\ell^2}\,, \qquad \beta = -\frac{b}{\ell^2}\,.
\end{equation}
The gravity dual of the `boosting' fluid on $\text{H}^2$ is thus given by the black hole solution
\eqref{boostingAdS4}, \eqref{A-boostingAdS4}\footnote{Strictly speaking, in order to prove this
rigorously, one would have to apply the map (4.1) of \cite{Bhattacharyya:2008mz}, and show that
this yields (up to second order in the boundary derivative expansion) the metric \eqref{boostingAdS4}.
We leave this for future work. In this context, note also that \cite{Bhattacharyya:2008mz} deals only
with the uncharged case. We are not aware of a magnetohydrodynamical generalization of the results
of \cite{Bhattacharyya:2008mz}.}.
Although the latter is contained in the general
Carter-Pleba\'nksi solution, it is in principle new, since its physical properties have not been
discussed in the literature so far. Note again the remarkable fact that the conformal transformation
\eqref{coord change conf bound mod}, \eqref{Weyl-boost} maps the boundary geometry
of \eqref{boostingAdS4} precisely to the region of $\mathbb{R}\times\text{H}^2$ in which
the fluid velocity does not exceed the speed of light.

\subsection{Black holes dual to mixed (parabolic) flow on the hyperbolic plane}

\begin{figure}
\centerline{
\includegraphics[scale=0.75]{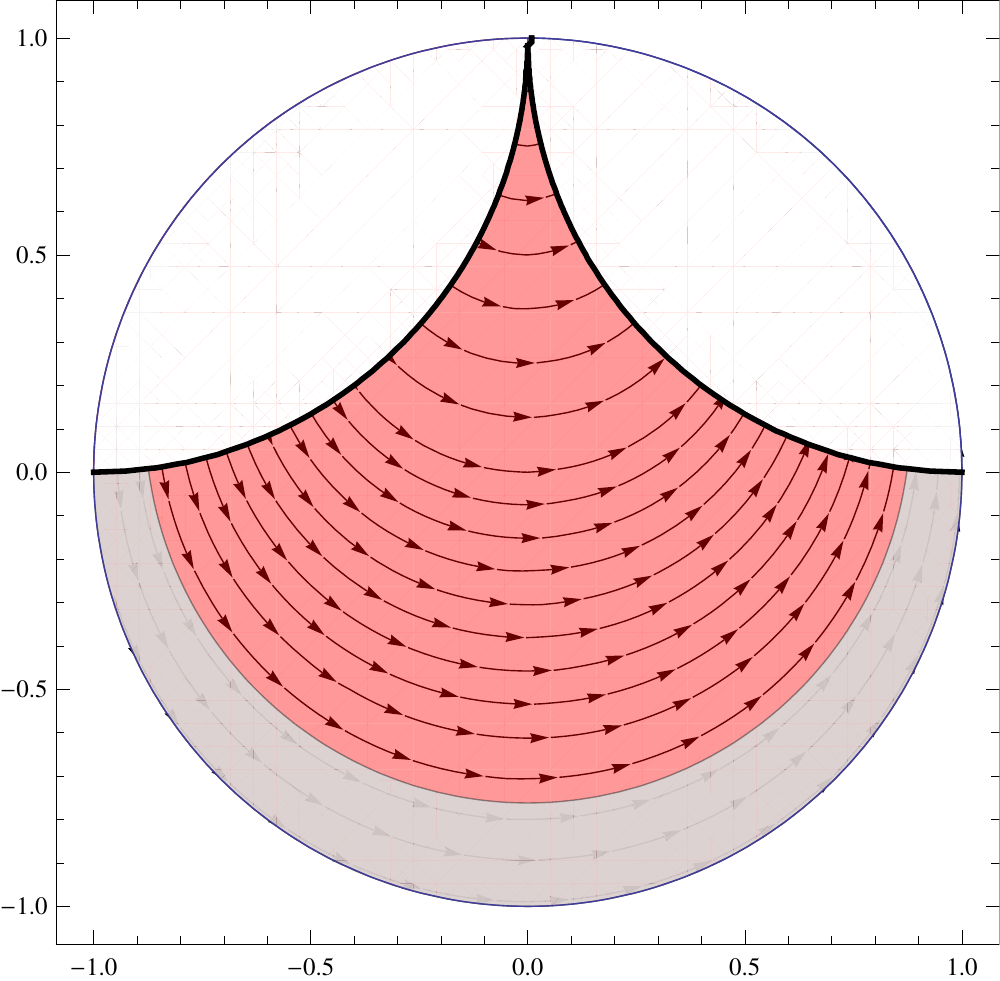}}
\caption{Poincar\'{e} disk compactified by identifications of the coordinate $B$; the two thick black lines
have to be identified. The grey area is the unphysical region where the fluid velocity exceeds the speed
of light. The fluid is located in the red region.}
\label{Grafico PD ident}
\end{figure}

Consider now the following choice for the parameters and coordinates of the Carter-Pleba\'nski
solution \eqref{Carter-Plebanski}:
\begin{displaymath}
\alpha = \mathrm{g}^2\,, \qquad l = 0\,, \qquad \epsilon = -1\,, \qquad q = r\,, \qquad p = aP\,,
\qquad \sigma = -\frac{\varphi}{a}\,,
\end{displaymath}
which leads to
\begin{equation} \label{gravdual-mixed}
\mathrm{d}s^2 = -\frac{\Delta_r}{\rho^2}\left(\mathrm{d}\tau + aP^2\mathrm{d}\varphi\right)^2
+ \rho^2\left(\frac{\mathrm{d}r^2}{\Delta_r} + \frac{\mathrm{d}P^2}{\Delta_P}\right) +
\frac{\Delta_P}{\rho^2}\left(a\mathrm{d}\tau - r^2\mathrm{d}\varphi\right)^2\,,
\end{equation}
\begin{equation}
A = -\frac{\mathrm{e}r}{\rho^2}\left(\mathrm{d}\tau+a P^2\mathrm{d}\varphi\right)
-\frac{\mathrm{g}P}{\rho^2}\left(a\mathrm{d}\tau-r^2\mathrm{d}\varphi\right)\,,
\end{equation}
where
\begin{displaymath}
\rho^2 = r^2+a^2P^2\,, \qquad \Delta_r = r^2\left(\frac{r^2}{\ell^2}-1\right) - 2mr + \mathrm{e}^2
+ \mathrm{g}^2\,, \qquad \Delta_P = P^2\left(1+\frac{a^2}{\ell^2}P^2\right)\,.
\end{displaymath}
The metric on the conformal boundary of \eqref{gravdual-mixed} reads
\begin{equation}
{\hat g} = -\mathrm{d}\tau^2+\frac{\ell^2}{\Delta_P}\mathrm{d}P^2+\ell^2P^2\mathrm{d}\varphi^2
-2aP^2\mathrm{d}\tau\mathrm{d}\varphi\,,
\end{equation}
and the holographic stress tensor takes the usual form
\begin{equation} \label{T-usual}
\hat{T}_{\mu\nu} = \frac{m}{8\pi\ell^2}\left({\hat g}_{\mu\nu}+3u_\mu u_\nu\right)\,,
\end{equation}
with 3-velocity $u=\partial_{\tau}$. Like in the previous cases, one can introduce ultrastatic
coordinates on the conformal boundary, defined by
\begin{equation} \label{ultrastat-coord-mixed}
T = \tau\,, \qquad A = \frac12\ln\frac{\Delta_P}{P^4}\,, \qquad B = \varphi-\frac{a\tau}{\ell^2}\,,
\end{equation}
where $A>\frac{1}{2}\ln\frac{a^2}{\ell^2}$ and $T,B\in\mathbb{R}$. After a subsequent
Weyl rescaling $\tilde{g}=\Omega^2{\hat g}$ with
\begin{equation}
\Omega = \frac{P}{\sqrt{\Delta_P}} = \sqrt{1-\frac{a^2}{\ell^2}e^{-2A}}\,, \label{Weyl-mixed}
\end{equation}
one gets the metric
\begin{equation}
\tilde{g} = -\mathrm{d}T^2+\ell^2(\mathrm{d}A^2+e^{-2A}\mathrm{d}B^2)\,.
\end{equation}
Thus we have shown that the boundary geometry of \eqref{gravdual-mixed} is conformal to the subset $A>\frac{1}{2}\ln\frac{a^2}{\ell^2}$ of the spacetime $\mathbb{R}\times\text{H}^2$.
After the coordinate transformation \eqref{ultrastat-coord-mixed} and the Weyl rescaling
\eqref{Weyl-mixed}, the 3-velocity $u$ becomes
\begin{equation}
\tilde{u} = \gamma(\partial_T-\frac{a}{\ell^2}\partial_B)\,,
\end{equation}
where
\begin{equation}
\gamma = \Omega^{-1} = \left(1-\frac{a^2}{\ell^2}e^{-2A}\right)^{-1/2}\,.
\end{equation}
The transformed energy-momentum tensor $\tilde T$ and the 3-velocity $\tilde u$ coincide with
the ones considered in section \ref{sec:mixed-flow}, if we identify
\begin{equation}
{\cal P}_0 = \frac{m}{8\pi\ell^2}\,, \qquad \beta = -\frac{a}{\ell^2}\,.
\end{equation}
The gravity dual of the mixed (parabolic) flow on $\text{H}^2$ is thus given by the black hole
solution \eqref{gravdual-mixed}. Again, the conformal transformation \eqref{ultrastat-coord-mixed},
\eqref{Weyl-mixed} maps the boundary of \eqref{gravdual-mixed} exactly to the region
$1-\beta^2\ell^2e^{-2A}>0$ where the fluid velocity is smaller than the speed of light.
If $\varphi$ is compactified, $B$ becomes also a compact coordinate. The flow in this case is
visualized in fig.~\ref{Grafico PD ident}.

\subsection{Black holes dual to rotating fluid on the Euclidean plane}

The family of black holes dual to a rotating fluid in Minkowski space $\mathbb{R}\times\text{E}^2$
is obtained by making the following choice for the parameters and coordinates of the Carter-Pleba\'nski
solution \eqref{Carter-Plebanski}:
\begin{displaymath}
\alpha = \mathrm{g}^2\,, \qquad l = 0\,, \qquad \epsilon = \frac{a^2}{\ell^2}\,, \qquad q = r\,,
\qquad p = aP\,, \qquad \sigma = -\frac{\varphi}{a}\,,
\end{displaymath}
which leads to
\begin{equation} \label{gravdual-rotE2}
\mathrm{d}s^2 =-\frac{\Delta_r}{\rho^2}\left(\mathrm{d}\tau+a P^2\mathrm{d}\varphi\right)^2 +
\rho^2\left(\frac{\mathrm{d}r^2}{\Delta_r} + \frac{\mathrm{d}P^2}{\Delta_P}\right) +
\frac{\Delta_P}{\rho^2}\left(a\mathrm{d}\tau - r^2\mathrm{d}\varphi\right)^2\,,
\end{equation}
\begin{equation}
A = -\frac{\mathrm{e}r}{\rho^2}\left(\mathrm{d}\tau + a P^2\mathrm{d}\varphi\right)
-\frac{\mathrm{g}P}{\rho^2}\left(a\mathrm{d}\tau - r^2\mathrm{d}\varphi\right)\,,
\end{equation}
where
\begin{displaymath}
\rho^2 = r^2+a^2P^2\,, \qquad \Delta_r = \frac{r^2}{\ell^2}(r^2+a^2)-2mr + \mathrm{e}^2
+ \mathrm{g}^2\,, \qquad \Delta_P = \frac{a^2}{\ell^2}P^2(P^2-1)\,,
\end{displaymath}
and $P>1$. In the uncharged case ($\mathrm{e}=\mathrm{g}=0$), the solution \eqref{gravdual-rotE2}
appeared in (C.10) of \cite{Mukhopadhyay:2013gja}.
Notice that, unlike in the previous cases, the Killing vector $\partial_\varphi$ becomes
timelike for large $r$. Hence, to avoid closed timelike curves, we shall not compactify $\varphi$.
Instead we replace the coordinates $\tau,\varphi$ with $T,\Phi$ defined by
\begin{equation}
T = \tau+a\varphi\,, \qquad \Phi=\frac{a\tau}{\ell^2}\,. \label{T-Phi-rotE2}
\end{equation}
Since $\partial_\Phi$ is spacelike everywhere outside the horizon (located at the largest root of
$\Delta_r$), we compactify $\Phi\sim\Phi+2\pi$. This choice is done in order that the conformal boundary has the topology of $\mathbb{R}$ times a disk, as we will
see shortly.

The boundary geometry of \eqref{gravdual-rotE2} is given by
\begin{equation}
{\hat g} = -\mathrm{d}\tau^2 + \frac{\ell^2}{\Delta_P}\mathrm{d}P^2 -a^2P^2\mathrm{d}\varphi^2
- 2aP^2\mathrm{d}\tau\mathrm{d}\varphi\,,
\end{equation}
and the holographic stress tensor has the usual form \eqref{T-usual}, where $u=\partial_\tau$.
Now consider the coordinate transformation \eqref{T-Phi-rotE2}, supplemented by
\begin{equation}
R=\frac{\ell^2}{a}\sqrt{1-P^{-2}}\,, \qquad 0 < R < \frac{\ell^2}{a}\,,
\end{equation}
and perform a Weyl rescaling $\tilde{g}=\Omega^2{\hat g}$ with
\begin{equation}
\Omega = P^{-1} = \sqrt{1-\frac{a^2 R^2}{\ell^4}}\,. \label{Weyl-rotE2}
\end{equation}
This yields the metric
\begin{equation}
\tilde{g} = -\mathrm{d}T^2+\mathrm{d}R^2+R^2\mathrm{d}\Phi^2\,.
\end{equation}
Thus we have shown that the conformal boundary is the subset $R<\ell^2/a$ of the spacetime $\mathbb{R}\times\text{E}^2$, i.e., the real line times a disk. After the coordinate change
to $(T,R,\Phi)$ and the conformal rescaling \eqref{Weyl-rotE2}, the energy momentum tensor becomes
\begin{equation} \label{stress-rotE2}
\tilde{T}^{\mu\nu} = \tilde{\cal P}\left(\tilde{g}^{\mu\nu}+3\tilde{u}^\mu\tilde{u}^\nu\right)\,,
\end{equation}
where
\begin{displaymath}
\tilde{\cal P} = \frac{m\gamma^3}{8\pi\ell^2}\,, \qquad \tilde{u} = \gamma\left(\partial_T +
\frac{a}{\ell^2}\partial_\Phi\right)\,, \qquad \gamma = \Omega^{-1} = \left(1-\frac{a^2 R^2}{\ell^4}
\right)^{-1/2}\,.
\end{displaymath}
The stress tensor \eqref{stress-rotE2} and the 3-velocity $\tilde u$ coincide with
the ones considered in section \ref{sec:rot-plane}, if we identify
\begin{equation}
{\cal P}_0 = \frac{m}{8\pi\ell^2}\,, \qquad \omega = \frac{a}{\ell^2}\,.
\end{equation}
The gravity dual of the rotating fluid on $\text{E}^2$ is thus given by the black hole
solution \eqref{gravdual-rotE2}, and the conformal transformation that we used here
maps the boundary geometry of \eqref{gravdual-rotE2} to a line times the disk
$R<\ell^2/a$, where the fluid flow is well-defined.

\subsection{Super-rotating hyperbolic black holes}

We saw in section \ref{Kerr-Newman-AdS4 BH} that the spherical KNAdS black hole is dual to a
rotating fluid on $\text{S}^2$ if the angular velocity of the latter is limited by $|\omega|\ell<1$, which
translates on the gravity side into $a^2<\ell^2$. For $|\omega|\ell>1$ the constraint $v^2<1$ restricts
the rotating fluid to the polar caps $|\omega|\ell\sin\theta<1$. It turns out that in this case the dual
black hole can be obtained from the KNAdS$_4$ metric \eqref{sph-hyp KNAdS4} with $k=1$ by
the analytical continuation $\theta\rightarrow i\theta$, which leads to
\begin{equation}
\mathrm{d}s^2 = -\frac{\Delta_r}{\Xi^2\rho^2}\left(\mathrm{d}t + a\sinh^2\theta\mathrm{d}\varphi
\right )^2 + \rho^2\left(\frac{\mathrm{d}r^2}{\Delta_r} + \frac{\mathrm{d}\theta^2}{\Delta_\theta}
\right ) + \frac{\Delta_\theta\sinh^2\theta}{\Xi^2\rho^2}\left(a\mathrm{d}t - (r^2+a^2)\mathrm{d}\varphi\right)^2\,, \label{hyp super KNAdS4}
\end{equation}
\begin{equation}
A = -\frac{\mathrm{e}r}{\Xi\rho^2}\left(\mathrm{d}t + a\sinh^2\theta\mathrm{d}\varphi\right)
- \frac{\mathrm{g}\cosh\theta}{\Xi\rho^2}\left(a\mathrm{d}t - (r^2+a^2)\mathrm{d}\varphi\right)\,,
\end{equation}
where
\begin{displaymath}
\rho^2 = r^2+a^2\cosh^2\theta\,, \quad
\Delta_r = (r^2+a^2)\left(1+\frac{r^2}{\ell^2}\right) - 2mr + \mathrm{e}^2 + \mathrm{g^2}\,, \quad
\Delta_\theta = \frac{a^2}{\ell^2}\cosh^2\theta - 1\,,
\end{displaymath}
and $\Xi=a^2/\ell^2-1$. Note that there is a lower bound on the rotation parameter $a$: Positivity
of $\Delta_\theta$ requires $a^2>\ell^2$, so that these black holes exist only above some minimum
amount of rotation and thus have no static limit.

The metric \eqref{hyp super KNAdS4} is again a special case of the Carter-Pleba\'nski family, obtained by
setting
\begin{displaymath}
\gamma = a^2+\mathrm{g}^2\,, \quad l = 0\,, \quad \epsilon = 1+\frac{a^2}{\ell^2}\,, \quad
\tau = \frac{t-a\varphi}{\Xi}\,, \quad q = r\,, \quad p = a\cosh\theta\,, \quad \sigma =
-\frac{\varphi}{a\Xi}\,.
\end{displaymath}
The metric on the conformal boundary of \eqref{hyp super KNAdS4} is given by
\begin{equation}\label{metric conf bound sup rotating bh}
{\hat g} = -\frac{\mathrm{d}t^2}{\Xi^2} + \frac{\ell^2}{\Delta_\theta}\mathrm{d}\theta^2 +
\frac{\ell^2}{\Xi}\sinh^2\theta\mathrm{d}\varphi^2 - \frac{2a}{\Xi^2}\sinh^2\theta\mathrm{d}t
\mathrm{d}\varphi\,.
\end{equation}
Now introduce new coordinates $\tau,\Theta,\Phi$ defined by
\begin{equation}\label{coord change conf bound suprot}
\tau = \frac{t}{\Xi}\,, \qquad \sin\Theta = \frac{\sinh\theta}{\sqrt{\Delta_\theta}}\,, \qquad
\Phi = \varphi-\frac{at}{\ell^2\Xi}\,,
\end{equation}
where $0<\Theta<\arcsin(\ell/|a|)$, and perform a Weyl rescaling $\tilde{g}=\Omega^2{\hat g}$
with
\begin{equation}
\Omega = \sqrt{\frac{\Xi}{\Delta_\theta}}\,. \label{Weyl-suprot}
\end{equation}
This gives
\begin{equation}
{\tilde g} = -\mathrm{d}\tau^2 + \ell^2(\mathrm{d}\Theta^2 + \sin^2\Theta\mathrm{d}\Phi^2)\,,
\end{equation}
and thus the conformal boundary of \eqref{hyp super KNAdS4} is, up to conformal transformations, the
polar cap $\Theta<\arcsin(\ell/|a|)$ of $\mathbb{R}\times \text{S}^2$.

After the conformal transformation \eqref{coord change conf bound suprot}, \eqref{Weyl-suprot}, the
holographic stress tensor associated to the spacetime \eqref{hyp super KNAdS4} becomes
\begin{equation}
\tilde{T}_{\mu\nu} = \tilde{\cal P}\left(\tilde{g}_{\mu\nu} + 3\tilde{u}_\mu \tilde{u}_\nu\right)\,,
\end{equation}
where
\begin{equation}
\tilde{\cal P} = \frac{m\gamma^3}{8\pi\ell^2}\,, \qquad \tilde{u} = \gamma\left(\partial_\tau -
\frac{a}{\ell^2}\partial_\Phi\right)\,, \qquad \gamma = \Omega^{-1} = \left(1 - \frac{a^2}{\ell^2}
\sin^2\Theta\right)^{-1/2}\,.
\end{equation}

$\tilde{T}$ is exactly the stress tensor \eqref{fluid on sphere stress tensor} of the stationary conformal
fluid on the 2-sphere with $|\omega|\ell>1$, if we identify
\begin{equation}
{\cal P}_0 = \frac{m}{8\pi\ell^2}\,, \qquad \omega = -\frac{a}{\ell^2}\,.
\end{equation}

\section{Final remarks} \label{fin-rem}

In this paper, we used hydrodynamics in order to make predictions on the possible types
of black holes in Einstein-Maxwell-AdS gravity. In particular, we classified the stationary equilibrium
flows on ultrastatic manifolds with spatial sections of constant curvature, and then used these
results to identify the dual black hole solutions. Although these are all contained in the
Carter-Pleba\'nski family, only a few of them have been studied in the literature before, so that
the major part is in principle new. The following table summarizes the results, relating to each
spacetime the corresponding dual fluid configuration.
\vspace{2 mm}

\noindent \centerline{
{\small{
    \begin{tabular}{|p{4.9cm}|p{2.8cm}||p{4.1cm}|p{1.4cm}|}
    \hline 
    Spacetime & Equ. & Fluid configuration & Section \\
    \hline \hline
    Spherical Kerr-Newman-AdS$_4$ & \eqref{sph-hyp KNAdS4}, with $k=1$   & Fluid in rigid
    rotation on the 2-sphere with $\omega < 1/\ell$ & \ref{stat conf fluid sph} \\ \hline
    Solution \eqref{gravdual-rotE2} & \eqref{gravdual-rotE2} &
    Fluid in rigid rotation on the Euclidean plane & \ref{sec:rot-plane} \\ \hline
    Hyperbolic KNAdS$_4$ & \eqref{sph-hyp KNAdS4}, with $k=-1$ &  Fluid in rigid
    rotation on the hyperbolic plane    & \ref{fluid rig rot hyp} \\ \hline
    Boosting AdS$_4$ black hole & \eqref{boostingAdS4} & Fluid
    translating on the hyperbolic plane & \ref{sec:transl-flow} \\ \hline
    Solution \eqref{gravdual-mixed} & \eqref{gravdual-mixed} & Mixed (parabolic) flow on
    the hyperbolic plane & \ref{sec:mixed-flow} \\ \hline
    Super-rotating hyperbolic black hole & \eqref{hyp super KNAdS4} & Fluid in rigid rotation on the
    2-sphere with $\omega > 1/\ell$ & \ref{stat conf fluid sph} \\ \hline
    \end{tabular}}} }

\vspace*{4mm}
It would be interesting to study more in detail the physics of these new black holes. Another possible
direction for future work is to repeat our analysis for hydrodynamics in four dimensions
(cf.~e.g.~\cite{Gubser:2010ui} for work in this direction) and to see if the dual metrics still
enjoy any sort of algebraic speciality.

Some remaining open questions concern for instance the boundary geometries of the
Carter-Pleba\'nski metric that have either no irrotational Killing field $\xi$ ($\Delta<0$ in
app.~\ref{app-CP}), or where $\xi$ is lightlike ($\Delta=0$). These cases include the black holes
with noncompact horizons but finite entropy constructed recently
in \cite{Gnecchi:2013mja,Klemm:2014rda}, as well as the cylindrical (or planar) solutions
of \cite{Klemm:1997ea}.
Although the boundary metric \eqref{conf-bdry-CP} is still conformally flat
for $\Delta\le 0$ (if the NUT charge vanishes), we were not able to find the coordinate transformation
that makes this manifest. However, the explicit form of this diffeomorphism would be needed in order
to quantitatively determine the hydrodynamic flow that is dual to these black holes.

Another intriguing point is the absence of a net Lorentz force acting on the charged fluid
on the boundary, as we saw in section \ref{dual-AdS-BH}. It would be very interesting to see
if this can be relaxed and, if so, what the holographic duals of such fluid configurations are.
For instance, one might ask which gravity dual corresponds to a charged fluid rotating in a plane, with
only a magnetic field orthogonal to that plane.

In section \ref{H2-to-plane} we saw that a conformal fluid in rigid rotation on hyperbolic space
looks completely different when transformed to the 2-sphere or the plane: There it becomes highly
dynamical, and takes the form of an expanding or contracting vortex. There is thus no need
to have dynamical spacetimes (which are notoriously difficult to construct) in order to build holographic
models of nonstationary (conformal) fluids.
This raises the question if bulk geometries of the type considered here can have applications
in a holographic description of the (dynamical) quark-gluon plasma produced in heavy-ion collisions,
cf.~\cite{McInnes:2013wba} for first attempts in this direction.
We hope to come back to some of these points in the future.

\acknowledgments

This work was partially supported by INFN. The authors would like to thank M.~M.~Caldarelli for
useful comments on the manuscript.

\appendix

\section{Notes on the Carter-Pleba\'nski metric} \label{app-CP}

In this section, we present a systematic classification of the possible types of black holes
contained in the Carter-Pleba\'nski family \eqref{Carter-Plebanski}, with a particular
emphasis on the geometries that can arise on the conformal boundary. The various cases are
distinguished by the number of real roots of the function $\mathsf{P}(p)$. This function must be
positive in order for the induced metric on the horizon to have the right signature.
We consider the case of vanishing NUT charge only, $l=0$, and define $\Gamma=\alpha-\mathrm{g}^2$,
such that $\mathsf{P}(p)$ in \eqref{PQ} boils down to
\begin{equation}
\mathsf{P}(p) = \frac{p^4}{\ell^2}-\epsilon p^2+\Gamma\,.
\end{equation}
Consider the discriminant $\Delta=\epsilon^2-4\Gamma/\ell^2$. For $\Delta\geq0$ one has
\begin{equation}
\mathsf{P}(p) = \frac{1}{\ell^2}(p^2-\alpha_+)(p^2-\alpha_-)\,, \label{P-alpha-pm}
\end{equation}
where $\alpha_{\pm}=\ell^2(\epsilon\pm\sqrt\Delta)/2$. We have then the following subcases:
\begin{enumerate}
\item If $\Gamma>0,\ \epsilon>2\sqrt\Gamma/\ell$, then $\Delta>0$, $\sqrt\Delta<\epsilon$,
so $\alpha_{\pm}>0$, and $\mathsf{P}$ has 4 real roots,
\begin{equation}
\mathsf{P}(p)=\frac{1}{\ell^2}(p-\sqrt\alpha_+)(p+\sqrt\alpha_+)(p-\sqrt\alpha_-)(p+\sqrt\alpha_-)\,.
\end{equation}
$\mathsf{P}$ is positive for $|p|>\sqrt\alpha_+$ or $|p|<\sqrt\alpha_-$.

In the latter region, use
the scaling symmetry \eqref{scaling-symm} to set $\alpha_+=\ell^2$ without loss of generality,
and define $a^2:=\alpha_-$. This gives the spherical KNAdS solution (\eqref{sph-hyp KNAdS4}
with $k=1$).

In the range $|p|>\sqrt\alpha_+$, use \eqref{scaling-symm} to set $\alpha_-=\ell^2$, and
define $a^2:=\alpha_+$, which leads to the super-rotating black hole \eqref{hyp super KNAdS4}.
\item If $\Gamma>0,\ \epsilon=2\sqrt\Gamma/\ell$, then $\Delta=0$, so $\alpha_{\pm}=\ell\sqrt\Gamma$, and
\begin{equation}
\mathsf{P}(p) = \frac{1}{\ell^2}\left(p-\sqrt{\ell\sqrt\Gamma}\right)^2\left(p+\sqrt{\ell\sqrt\Gamma}
\right)^2\,.
\end{equation}
$\mathsf{P}$ is positive for $p\neq\pm\sqrt{\ell\sqrt\Gamma}$. By virtue of \eqref{scaling-symm}
one can always take $\epsilon=2$, i.e., $\sqrt{\ell\sqrt{\Gamma}}=\ell$.
Then, for $|p|<\ell$, we get the black holes that have a noncompact
horizon with finite area, constructed recently in \cite{Gnecchi:2013mja,Klemm:2014rda}. For
$|p|>\ell$ one obtains new solutions that have not been discussed in the literature so far.
\item If $\Gamma>0,\ -2\sqrt\Gamma/\ell<\epsilon<2\sqrt\Gamma/\ell$, then $\Delta<0$, so $\mathsf{P}$ has no real roots and is always positive. These solutions are new, except the case
$\epsilon=0$, which corresponds (with the definition $a^2:=\Gamma$) to the cylindrical black holes
found in \cite{Klemm:1997ea}.
\item If $\Gamma>0,\ \epsilon=-2\sqrt\Gamma/\ell$, then $\Delta=0$, so $\alpha_{\pm}=-\ell\sqrt\Gamma$. $\mathsf{P}$ has no real roots and is always positive,
\begin{equation}
\mathsf{P}(p)=\frac{1}{\ell^2}(p^2+\ell\sqrt\Gamma)^2\,.
\end{equation}
Also this case has not been considered in the literature yet.
\item If $\Gamma>0,\ \epsilon<-2\sqrt\Gamma/\ell$, then $\Delta>0$, $\epsilon<-\sqrt\Delta$, so $\alpha_{\pm}<0$. $\mathsf{P}$ has no real roots and is given by \eqref{P-alpha-pm}. It is easy to
see that one can always use \eqref{scaling-symm} to set $\epsilon=-1-\Gamma/\ell^2$. If we define
$b^2:=\Gamma$, we obtain the boosting AdS$_4$ back holes \eqref{boostingAdS4}.
\item If $\Gamma=0,\ \epsilon>0$, then $\Delta>0$, $\alpha_+=\epsilon\ell^2$, $\alpha_-=0$, and
\begin{equation}
\mathsf{P}(p)=\frac{p^2}{\ell^2}(p-\ell\sqrt\epsilon)(p+\ell\sqrt\epsilon)\,,
\end{equation}
which is positive for $|p|>\ell\sqrt\epsilon$. This case yields the solution \eqref{gravdual-rotE2},
dual to a rotating fluid on $\mathbb{R}\times\text{E}^2$, with rotation parameter $a$ given by
$\epsilon=a^2/\ell^2$.
\item If $\Gamma=0,\ \epsilon=0$, then $\Delta=0$, $\alpha_+=\alpha_-=0$, and
$\mathsf{P}(p)=p^4/\ell^2$. This is again a hitherto undiscussed geometry.
\item If $\Gamma=0,\ \epsilon<0$, then $\Delta>0$, $\alpha_+=0$, $\alpha_-=\epsilon\ell^2$, and
\begin{equation}
\mathsf{P}(p)=\frac{p^2}{\ell^2}(p^2-\epsilon\ell^2)\,,
\end{equation}
which is positive for $p\neq 0$. By means of \eqref{scaling-symm} one can scale $\epsilon=-1$,
and gets the solution \eqref{gravdual-mixed}, dual to a mixed flow on $\mathbb{R}\times\text{H}^2$.
\item If $\Gamma<0$, then $\Delta>0$, $\sqrt\Delta>|\epsilon|$, $\alpha_+>0$, $\alpha_-<0$, and
\begin{equation}
\mathsf{P}(p)=\frac{1}{\ell^2}(p-\sqrt{\alpha_+})(p+\sqrt{\alpha_+})(p^2-\alpha_-)\,.
\end{equation}
$\mathsf{P}$ is positive for $|p|>\sqrt{\alpha_+}$. Use \eqref{scaling-symm} to set $\alpha_-=-\ell^2$
and define $a$ by $a^2=\alpha_+$. This leads to the hyperbolic KNAdS black hole, i.e.,
\eqref{sph-hyp KNAdS4} with $k=-1$.
\end{enumerate}

\subsection{The static Killing fields of the conformal boundary} \label{stat-kill-bdry}

The metric $\hat g$ on the conformal boundary of the Carter-Pleba\'nski family is given by
\eqref{conf-bdry-CP}. The only Killing fields $\xi$ of $\hat g$ are linear combinations of $\partial_\tau$
and $\partial_\sigma$, i.e., $\xi=A\partial_\tau+B\partial_\sigma$, and the orthogonal distribution
of $\xi$ is generated by the fields $f\partial_\tau+\partial_\sigma$ and $\partial_p$, where
\begin{equation}
f=\frac{A p^2+B(\ell^2\mathsf{P}(p)-p^4)}{A-B p^2}\,.
\end{equation}
$\xi$ is irrotational if and only if $\Delta\geq0$ and $A=\alpha_{\pm}B$. With this choice, the function
$f$ reduces to $f_\pm=\alpha_\mp$. To see this, consider the orthogonal distribution of $\xi$,
which is involutive if and only if $[f\partial_\tau+\partial_\sigma,\partial_p]=-\partial_p f\partial_\tau$
belongs to it, which happens when $\partial_pf$ vanishes, i.e.\ when
$A^2-\epsilon\ell^2 AB+B^2\ell^2\Gamma=0$. This equation has solutions $A$ for $\Delta\ge 0$;
these are $A_\pm=\ell^2B(\epsilon\pm\sqrt{\Delta})/2=\alpha_\pm B$. Plugging
$A_\pm$ into $f$ yields $f_\pm=\alpha_\mp$.
The only irrotational Killing fields are thus multiples of
\begin{equation}
\xi_\pm=\alpha_\pm\partial_\tau+\partial_\sigma\,,
\end{equation}
and the orthogonal distribution of $\xi_\pm$ is generated by $\xi_\mp$ and $\partial_p$.

Now introduce, for $\Delta\geq 0$, the functions
\begin{equation}
\Psi_\pm(p) = {\hat g}(\xi_\pm,\xi_\pm)=\pm\ell^2\sqrt\Delta(p^2-\alpha_\pm )\,.
\end{equation}
We have then:
\begin{itemize}
 \item For $\Delta<0$ (case 3) there are no irrotational Killing fields.
 \item For $\Delta=0$ (cases 2,4,7), $\xi_+=\xi_-$ is lightlike, so there is no static Killing field.
 \item For $\Gamma>0$ and $\epsilon>2\sqrt\Gamma/\ell$ (case 1), one has $\alpha_\pm>0$, so
$\xi_+$ is timelike for $|p|<\sqrt{\alpha_+}$ and spacelike for $|p|>\sqrt\alpha_+$, while
$\xi_-$ is timelike for $|p|>\sqrt{\alpha_-}$ and spacelike for $|p|<\sqrt{\alpha_-}$.
 \item If $\Gamma>0$ and $\epsilon<-2\sqrt\Gamma/\ell$ (case 5), then $\alpha_\pm<0$, and thus
$\xi_+$ is always spacelike, whereas $\xi_-$ is always timelike.
 \item If $\Gamma=0$ and $\epsilon>0$ (case 6), then $\alpha_+>0$, $\alpha_-=0$, hence
$\xi_+$ is timelike for $|p|<\sqrt{\alpha_+}$ and spacelike for $|p|>\sqrt{\alpha_+}$, while
$\xi_-$ is timelike for $p\neq 0$ and never spacelike.
 \item When $\Gamma=0$ and $\epsilon<0$ (case 8), then $\alpha_+=0$, $\alpha_-<0$, so
$\xi_+$ is spacelike for $p\neq 0$ and never timelike, whereas $\xi_-$ is always timelike.
 \item For $\Gamma<0$ (case 9), we have $\alpha_+>0$, $\alpha_-<0$, and thus
$\xi_+$ is timelike for $|p|<\sqrt{\alpha_+}$ and spacelike for $|p|>\sqrt{\alpha_+}$, while
$\xi_-$ is always timelike.
\end{itemize}
In each case with $\Delta>0$, in the regions where $\mathsf{P}(p)>0$ either $\xi_+$ is spacelike and
$\xi_-$ is timelike or vice versa. $\xi_\pm$ do not change their causal character inside these regions,
and therefore the spacetime is static. Moreover, $\xi_+,\xi_-,\partial_p$ form an orthogonal frame.
Now introduce coordinates $\tau_\pm$ such that $\partial_{\tau_\pm}=\xi_\pm$, given by
\begin{equation}\label{static coordinates on CP boundary}
\tau_\pm = \pm\frac{1}{\ell^2\sqrt\Delta}(\tau-\alpha_\mp\sigma)\,.
\end{equation}
In these coordinates the boundary metric \eqref{conf-bdry-CP} reads
\begin{equation}
{\hat g} = \Psi_+(p)\mathrm{d}\tau_+^2+\Psi_-(p)\mathrm{d}\tau_-^2+\frac{\ell^2}{\mathsf{P}(p)}\mathrm{d}p^2\,.
\end{equation}
If $\xi_+$ is timelike ($\Psi_+<0$), then we rescale $\hat g$ with $\Omega^2=-\kappa/\Psi_+(p)$
(where $\kappa\in\mathbb{R}_+$ has been introduced for later convenience) to get the ultrastatic metric
\begin{equation}
\tilde g = -\kappa\mathrm{d}\tau_+^2-\frac{\kappa}{\Psi_+(p)}\left(\frac{\ell^2}{\mathsf{P}(p)}
\mathrm{d}p^2+\Psi_-(p)\mathrm{d}\tau_-^2\right)\,. \label{metr:ultra:+}
\end{equation}
Note that the sections of constant $\tau_+$ have constant scalar curvature $R=2\alpha_+\Delta/\kappa$.
If $\xi_-$ is timelike ($\Psi_-<0$), then we rescale $\hat g$ with $\Omega^2=-\kappa/\Psi_-(p)$ to
get the ultrastatic metric
\begin{equation}
\tilde g = -\kappa\mathrm{d}\tau_-^2-\frac{\kappa}{\Psi_-(p)}\left(\frac{\ell^2}{\mathsf{P}(p)}
\mathrm{d}p^2+\Psi_+(p)\mathrm{d}\tau_+^2\right)\,, \label{metr:ultra:-}
\end{equation}
whose $\tau_-=\text{constant}$ sections have scalar curvature $R=2\alpha_-\Delta/\kappa$.

We have thus shown that in the cases 1,5,6,8,9 the spacetime has one static Killing field, and is conformal
to an ultrastatic manifold with spatial sections of constant curvature.
In what follows, we shall consider each of these cases separately, and show that they correspond
precisely to the equilibrium flows considered in section \ref{stat flow ultrastatic st}.

\begin{itemize}
\item Case 1, region $p\in(-\sqrt{\alpha_-},\sqrt{\alpha_-})$

Consider case 1 ($\Gamma>0$, $\epsilon>2\sqrt\Gamma/\ell$). Take the region
$p\in(-\sqrt{\alpha_-},\sqrt{\alpha_-})$, where $\mathsf{P}(p)>0$, $\Psi_+(p)<0$, $\Psi_-(p)>0$,
and thus the boundary metric is conformal to \eqref{metr:ultra:+}. If we choose
$\kappa=\ell^2\alpha_+\Delta$ (which is positive), the sections of constant $\tau_+$ have scalar
curvature $R=2/\ell^2$. Now introduce new coordinates $(T,\Theta,\Phi)$ defined by
\begin{equation}\label{spherical coordinates in case 1.1}
T = \ell\sqrt{\alpha_+\Delta}\tau_+\,, \qquad
\cos\Theta = \sqrt{\frac{\alpha_+ - \alpha_-}{\alpha_-(\alpha_+-p^2)}}p\,, \qquad
\Phi=-\sqrt{\alpha_-\Delta}\tau_-\,,
\end{equation}
where $\Theta$ ranges in $(0,\pi)$. Then \eqref{metr:ultra:+} simplifies to
\begin{equation} \label{g-tilde:case1.1}
\tilde g = -\mathrm{d}T^2+\ell^2\left(\mathrm{d}\Theta^2+\sin^2\Theta\mathrm{d}\Phi^2\right)\,.
\end{equation}
In section \ref{dual-AdS-BH} it was found that the 3-velocity of the fluid dual to the Carter-Plebanski
geometry is given by $u=\partial_\tau$. In the coordinates \eqref{spherical coordinates in case 1.1}
and after the conformal rescaling with $\Omega^2=-\kappa/\Psi_+(p)$, this becomes
\begin{equation} \label{u-tilde:case1.1}
\tilde{u}=\frac{1}{\sqrt{1-\omega^2\ell^2\sin^2\Theta}}\left(\partial_T+\omega\partial_\Phi\right)\,,
\end{equation}
with $\omega=\sqrt{\frac{\alpha_-}{\alpha_+}}\frac{1}{\ell}$. Notice that $\omega\in(0,\frac{1}{\ell})$.
This is precisely the flow on $\mathbb{R}\times\text{S}^2$ considered in section \ref{stat conf fluid sph}.

\item Case 1, region $p\in(-\infty,-\sqrt{\alpha_+})\cup(\sqrt{\alpha_+},+\infty)$

Consider still case 1, but this time take the region $|p|>\sqrt{\alpha_+}$, where $\mathsf{P}(p)>0$,
$\Psi_+(p)>0$, $\Psi_-(p)<0$, and thus the boundary metric is conformal to \eqref{metr:ultra:-}.
If we choose $\kappa=\ell^2\alpha_-\Delta$ (which is positive), the scalar curvature of the
constant $\tau_-$ sections becomes $R=2/\ell^2$. Now introduce new coordinates $(T,\Theta,\Phi)$
according to
\begin{equation}\label{spherical coordinates in case 1.2}
T = -\ell\sqrt{\alpha_-\Delta}\tau_-\,, \qquad
\sin\Theta = \sqrt{\frac{\alpha_-(p^2-\alpha_+)}{\alpha_+(p^2-\alpha_-)}}\,, \qquad
\Phi=\sqrt{\alpha_+\Delta}\tau_+\,,
\end{equation}
where now $\Theta$ ranges in $(0,\arcsin\sqrt{\frac{\alpha_-}{\alpha_+}})$ when $p\in(\sqrt{\alpha_+},+\infty)$ and $(\pi-\arcsin\sqrt{\frac{\alpha_-}{\alpha_+}},\pi)$ when $p\in(-\infty,-\sqrt{\alpha_+})$.
Then the metric \eqref{metr:ultra:-} becomes again \eqref{g-tilde:case1.1},
and the 3-velocity $u=\partial_\tau$ of the fluid is still transformed into \eqref{u-tilde:case1.1}, but
this time with $\omega=\sqrt{\frac{\alpha_+}{\alpha_-}}\frac{1}{\ell}$, which satisfies
$\omega>1/\ell$. Moreover, $\Theta$ is now restricted to the polar caps $\omega\ell\sin\Theta<1$.
This is again the flow on $\mathbb{R}\times\text{S}^2$ considered in section \ref{stat conf fluid sph},
but with $\omega>1/\ell$.

\item Case 5

In this case ($\Gamma>0$, $\epsilon<-2\sqrt\Gamma/\ell$) we have, for each $p\in\mathbb{R}$, $\mathsf{P}(p)>0$, $\Psi_+(p)>0$, $\Psi_-(p)<0$, hence the boundary metric is conformal to
\eqref{metr:ultra:-}. If we choose $\kappa=-\ell^2\alpha_-\Delta$ (which is positive), the sections
of constant $\tau_-$ have scalar curvature $R=-2/\ell^2$. After the coordinate change
\begin{equation}\label{hyperbolic coordinates in case 5}
T = -\ell\sqrt{-\alpha_-\Delta}\tau_-\,, \qquad
\sinh\Theta = \sqrt{\frac{\alpha_+ - \alpha_-}{-\alpha_+(p^2-\alpha_-)}}p\,, \qquad
\Phi = \sqrt{-\alpha_+\Delta}\tau_+\,,
\end{equation}
where $|\Theta|<\text{arcosh}\sqrt{\frac{\alpha_-}{\alpha_+}}$, the metric \eqref{metr:ultra:-}
boils down to
\begin{equation}
\tilde g = -\mathrm{d}T^2+\ell^2\left(\mathrm{d}\Theta^2+\cosh^2\Theta\mathrm{d}\Phi^2\right)\,,
\end{equation}
while the fluid velocity becomes
\begin{equation}
\tilde{u} = \frac{1}{\sqrt{1-\beta^2\ell^2\cosh^2\Theta}}\left(\partial_T+\beta\partial_\Phi\right)\,,
\end{equation}
with $\beta=\sqrt{\frac{\alpha_+}{\alpha_-}}\frac{1}{\ell}$. Notice that $\beta\in(0,1/\ell)$ and $\beta\ell\cosh\Theta<1$. This is the purely translational flow on $\mathbb{R}\times\text{H}^2$
of section \ref{sec:transl-flow}\footnote{Set $X=\cosh\Theta\sinh\Phi$, $Y=\sinh\Theta$ to
compare with section \ref{sec:transl-flow}.}.

\item Case 6

In this case ($\Gamma=0$, $\epsilon>0$), $\mathsf{P}(p)$ is positive for $|p|>\sqrt{\alpha_+}$, where
$\Psi_+(p)>0$, $\Psi_-(p)<0$. The boundary metric is thus conformal to \eqref{metr:ultra:-},
and (since $\alpha_-=0$) the scalar curvature of the constant $\tau_-$ sections vanishes.
Now put $\kappa=\ell^4\Delta$ and introduce new coordinates $(T,R,\Phi)$ defined by
\begin{equation}\label{polar coordinates in case 6}
T = -\ell^2\sqrt{\Delta}\tau_-\,, \qquad
R = \frac{\ell^2}{\sqrt{\alpha_+}}\sqrt{1-\frac{\alpha_+}{p^2}}\,, \qquad
\Phi=\frac{\sqrt{\alpha_+^3}}{\ell^2}\tau_+\,,
\end{equation}
where $0<R<\ell/\sqrt{\epsilon}$. Then \eqref{metr:ultra:-} turns into
\begin{equation}
\tilde g = -\mathrm{d}T^2+\mathrm{d}R^2+R^2\mathrm{d}\Phi^2\,,
\end{equation}
and the 3-velocity of the fluid becomes
\begin{equation}
\tilde{u} = \frac{1}{\sqrt{1-\omega^2 R^2}}\left(\partial_T+\omega\partial_\Phi\right)\,,
\end{equation}
with $\omega=\sqrt{\alpha_+}/\ell^2$. Notice that $R<1/\omega$. This is the rigidly
rotating fluid on Minkowski space considered in \ref{sec:rot-plane}.

\item Case 8

Here we have $\Gamma=0$, $\epsilon<0$, and $\mathsf{P}(p)>0$ for $p\neq 0$. Moreover,
$\Psi_+(p)>0$ and $\Psi_-(p)<0$, and thus the boundary metric is conformal to \eqref{metr:ultra:-}.
If we choose $\kappa=-\ell^2\alpha_-\Delta$ (which is positive), the scalar curvature of the sections
$\tau_-=\text{constant}$ becomes $R=-2/\ell^2$.
Now introduce new coordinates $(T,A,B)$ according to
\begin{equation}\label{special coordinates in case 8.1}
T = -\ell\sqrt{-\alpha_-\Delta}\tau_-\,, \qquad
A = \frac{1}{2}\log\left(1-\frac{\alpha_-}{p^2}\right)+\ln\ell\beta\,, \qquad
B = \ell\beta\sqrt{-\alpha_-\Delta}\tau_+\,,
\end{equation}
where we have introduced an arbitrary parameter $\beta>0$, which can be chosen as $\beta=1/\ell$
without loss of generality. Note that $\ln\ell\beta<A<\infty$. This casts \eqref{metr:ultra:-} into
the form
\begin{equation}
\tilde g = -\mathrm{d}T^2+\ell^2(\mathrm{d}A^2+e^{-2A}\mathrm{d}B^2)\,,
\end{equation}
while the 3-velocity $u=\partial_\tau$, after the conformal rescaling $\tilde g=\Omega^2\hat g$,
becomes
\begin{equation}
\tilde{u} = \frac{1}{\sqrt{1-\beta^2\ell^2 e^{-2A}}}\left(\partial_T+\beta\partial_B\right)\,.
\end{equation}
This corresponds to the mixed (parabolic) flow on $\mathbb{R}\times\text{H}^2$ of section
\ref{sec:mixed-flow}.

\item Case 9

The last case is $\Gamma<0$. The polynomial $\mathsf{P}(p)$ is positive for $p>\sqrt{\alpha_+}$,
where $\Psi_+(p)>0$, $\Psi_-(p)<0$. Therefore the boundary metric is conformal to \eqref{metr:ultra:-}.
If we choose $\kappa=-\ell^2\alpha_-\Delta$ (which is positive), the constant $\tau_-$ sections have
scalar curvature $R=-2/\ell^2$. After the coordinate change
\begin{equation}\label{hyperbolic coordinates in case 9.1}
T = -\ell\sqrt{-\alpha_-\Delta}\tau_-\,, \qquad
\sinh\Theta = \sqrt{\frac{-\alpha_-(p^2-\alpha_+)}{\alpha_+(p^2-\alpha_-)}}\,, \qquad
\Phi=\sqrt{\alpha_+\Delta}\tau_+\,,
\end{equation}
where $\Theta$ ranges in $(0,\text{arcsinh}\sqrt{-\alpha_-/\alpha_+})$, the metric \eqref{metr:ultra:-}
boils down to
\begin{equation}
\tilde g = -\mathrm{d}T^2+\ell^2\left(\mathrm{d}\Theta^2+\sinh^2\Theta\mathrm{d}\Phi^2\right)\,,
\end{equation}
and the 3-velocity of the fluid is
\begin{equation}
\tilde{u} = \frac{1}{\sqrt{1-\omega^2\ell^2\sinh^2\Theta}}\left(\partial_T+\omega\partial_\Phi\right)\,,
\end{equation}
with $\omega=\sqrt{-\alpha_+/\alpha_-}/\ell$. Notice that $\omega\ell\sinh\Theta<1$.
This corresponds to the rigidly rotating fluid on $\mathbb{R}\times\text{H}^2$, considered in
\ref{fluid rig rot hyp}.
\end{itemize}
Note that in all cases where the positivity region of $\mathsf{P}(p)$ consists of two disconnected
parts (1b,6,8,9), the corresponding coordinate transformations map both the branch where $p$ is
positive and the one where $p$ is negative to the same spacetime. (In case 1b up to isometries,
since the region $p<-\sqrt{\alpha_+}$ maps to the lower polar cap, while $p>\sqrt{\alpha_+}$
maps to the upper polar cap).

\section{Proof of propositions} \label{app-proof}

{\bf Prop.~\ref{prop-killing}:}

Equ.~\eqref{u=gamma(1,v)} implies that
\begin{equation}\label{theta}
\nabla_t u^\mu=0\,,\ \ \ \ \ \nabla_\mu u^t=\partial_\mu u^t\,,\ \ \ \ \ \nabla_i u^j=v^j\partial_i\gamma+\gamma\bar{\nabla}_i v^j\,, \ \ \ \ \
\vartheta = v^i\partial_i\gamma+\gamma\bar{\nabla}_i v^i\,,
\end{equation}
where $\bar{\nabla}$ denotes the Levi-Civita connection of $(\Sigma,\bar{g})$. Moreover 
\begin{equation}\label{Dgamma}
\partial_i\gamma=\gamma^3 v_j\bar{\nabla}_i v^j\,.
\end{equation}
These expressions can be used in \eqref{a-theta-sigma} to compute $\sigma^{\mu\nu}$, with the
result
\begin{flalign}
&\sigma^{tt} =
\frac{\gamma^2}{d-1}(d-1-v^2)v^i\partial_i \gamma - \frac{v^2\gamma^3}{d-1}\bar{\nabla}_iv^i\,,
\label{sigma^tt}\\
&\sigma^{ti} =
\frac{d-2}{d-1}\gamma^2 v^iv^j\partial_j\gamma+\frac{1}{2}\bar{g}^{ij}\partial_j\gamma+\gamma^3\left(\frac{1}{2}v^j\bar{\nabla}_j v^i-\frac{1}{d-1}v^i\bar{\nabla}_j v^j \right)\label{sigma^ti}\,,\\
&\sigma^{ij} =
\left(\frac{1}{2}\bar{g}^{ik}v^j+\frac{1}{2}\bar{g}^{jk}v^i-\frac{1}{d-1}\bar{g}^{ij}v^k \right )\partial_k\gamma+
\gamma\left(\frac{1}{2}\bar{\nabla}^iv^j+\frac{1}{2}\bar{\nabla}^jv^i-\frac{1}{d-1}\bar{g}^{ij}\bar{\nabla}_k v^k  \right )\nonumber\\
&\ \ \ \ \ 
+\frac{d-2}{d-1}\gamma^2 v^iv^jv^k\partial_k\gamma+
\gamma^3\left(\frac{1}{2}v^iv^k\bar{\nabla}_k v^j+\frac{1}{2}v^jv^k\bar{\nabla}_k v^i-\frac{1}{d-1}v^iv^j\bar{\nabla}_k v^k \right)\,.\label{sigma^ij}
\end{flalign}
Putting together eqns.~\eqref{sigma^ti} and \eqref{sigma^ij} we find that
\begin{flalign}
&\sigma^{ij}=
v^i\sigma^{tj}+v^j\sigma^{ti}-\frac{d-2}{d-1}\gamma^2 v^iv^jv^k\partial_k\gamma+
\frac{1}{d-1}\gamma^3 v^iv^j\bar{\nabla}_k v^k&\nonumber\\
&\ \ \ \ \ -\frac{1}{d-1}\bar{g}^{ij}(v^k\partial_k\gamma+\gamma\bar{\nabla}_k v^k)+\frac{\gamma}{2}(\bar{\nabla}^iv^j+\bar{\nabla}^jv^i)\,.\label{sigma^ij (2)}
\end{flalign}
Now let $\vartheta=0$ and $\sigma^{\mu\nu}=0$. From \eqref{sigma^tt} one gets
\begin{equation}\label{vDgamma}
v^i\partial_i\gamma=\frac{\gamma v^2}{d-1-v^2}\bar{\nabla}_iv^i\,.
\end{equation}
Using the last equ.~of \eqref{theta}, we obtain then
\begin{equation}
0 = \vartheta = v^i\partial_i\gamma +\gamma\bar{\nabla}_iv^i =
\gamma\frac{d-1}{d-1-v^2}\bar{\nabla}_iv^i\,,
\end{equation}
and thus
\begin{equation}\label{Div v=0}
\bar{\nabla}_iv^i=0\,.
\end{equation}
Plugging \eqref{vDgamma} into \eqref{sigma^ij (2)} yields
\begin{displaymath}
0=\sigma^{ij}=
\frac{\gamma\bar{\nabla}_k v^k}{(d-1)(d-1-v^2)}\left((d-1)v^iv^j-\gamma^2 v^2 v^iv^j-(d-1)\bar{g}^{ij} \right )+\frac{\gamma}{2}(\bar{\nabla}^iv^j+\bar{\nabla}^jv^i)\,,
\end{displaymath}
and hence, by \eqref{Div v=0},
\begin{equation}\label{v Killing}
\bar{\nabla}^iv^j+\bar{\nabla}^jv^i=0\,,
\end{equation}
i.e., $v$ is Killing.

Viceversa, suppose that $v$ is a Killing field: Taking the trace of \eqref{v Killing} gives \eqref{Div v=0}; moreover, eqns.~\eqref{v Killing} and \eqref{Dgamma} give
\begin{equation}\label{vDgamma=0}
v^i\partial_i\gamma=\gamma^3v^iv^j\bar{\nabla}_iv_j=0\,,
\end{equation}
so that $\vartheta=0$ by equ.~\eqref{theta}.
Now \eqref{sigma^tt} leads to $\sigma^{tt}=0$, while equ.~\eqref{sigma^ti} becomes, using
\eqref{vDgamma}, \eqref{Div v=0} and \eqref{Dgamma},
\begin{equation}
\sigma^{ti}=\frac{1}{2}\bar{g}^{ij}\partial_j\gamma+\frac{\gamma^3}{2}v^j\bar{\nabla}_j v^i=\frac{\gamma^3}{2}\bar{g}^{ij}v^k(\bar{\nabla}_j v_k+\bar{\nabla}_k v_j)=0\,.
\end{equation}
Finally, using these results in \eqref{sigma^ij (2)} we find $\sigma^{ij}=0$, which completes the proof.\\

{\bf Prop.~\ref{prop-NS}:}

Since $\vartheta=0$ and $\partial_t {\cal P}=0$, we have
\begin{equation}
\nabla_\mu T^{\mu\nu}=\partial_i{\cal P}(d\, u^i u^\nu+g^{i\nu})+{\cal P}d\,u^\mu\nabla_\mu u^\nu\,.
\end{equation}
Using eqns.~\eqref{u=gamma(1,v)}, \eqref{theta}, \eqref{vDgamma=0} one gets
\begin{displaymath}
u^\mu\nabla_\mu u^t=u^i\partial_i u^t=\gamma v^i\partial_i\gamma=0\,, \qquad
u^\mu\nabla_\mu u^j=u^i\nabla_i u^j=\gamma v^i(v^j\partial_i\gamma+\gamma\bar{\nabla}_iv^j)
= \gamma^2v^i\bar{\nabla}_iv^j\,,
\end{displaymath}
and thus
\begin{displaymath}
\nabla_\mu T^{\mu t}=d\,\gamma^2 v^i\partial_i {\cal P}\,, \qquad
\nabla_\mu T^{\mu j}=d\,\gamma^2 v^i v^j \partial_i {\cal P} + \partial^j {\cal P}+ {\cal P}d\gamma^2
v^i\bar{\nabla}_i v^j\,.
\end{displaymath}
The vanishing of these two expressions is equivalent to\footnote{Contracting \eqref{equiv-equ} with $v^j$
yields $v^j\partial_j {\cal P}=0$ by \eqref{v Killing}.}
\begin{equation} \label{equiv-equ}
\partial_j {\cal P} + {\cal P}d\gamma^2 v^i\bar{\nabla}_i v_j=0\,,
\end{equation}
which can be rewritten as\footnote{Use $\partial_j\gamma=\gamma^3 v^i\bar{\nabla}_j v_i=
-\gamma^3 v^i\bar{\nabla}_i v_j$, which follows from \eqref{Dgamma} and \eqref{v Killing}.}
$\partial_i\ln{\cal P}=d\partial_i\ln\gamma$.\\

{\bf Prop.~\ref{prop-heat-flux}:}

Using \eqref{u=gamma(1,v)} we get
\begin{displaymath}
a_t = -a^t = -\gamma v^i\partial_i\gamma\,, \qquad a_i = \bar{g}_{ij}a^j = \gamma v^k
(v_i\partial_k\gamma + \gamma\bar{\nabla}_kv_i)\,.
\end{displaymath}
Owing to $\partial_i\gamma=\gamma^3v_j\bar{\nabla}_iv^j$ one has moreover
\begin{displaymath}
P^{t\nu}a_\nu = \gamma^4 v^i v^j\bar{\nabla}_i v_j\,, \qquad P^{i\nu}a_\nu = \gamma^2 v^k
\bar{\nabla}_k v^i + \gamma v^iv^k\partial_k\gamma\,.
\end{displaymath}
The components of the heat flux in \eqref{subl-currents} become thus
\begin{flalign}
&q^t = -\kappa\left((\gamma^2-1)\partial_t{\cal T}+\gamma^2 v^i\partial_i{\cal T}+{\cal T}\gamma^4
v^i v^j\bar{\nabla}_iv_j\right)\,, \nonumber \\
&q^i = -\kappa\left(\gamma^2v^i\partial_t{\cal T}+(\bar{g}^{ij}+\gamma^2 v^iv^j)
\partial_j{\cal T}+{\cal T}(\gamma^2v^k\bar{\nabla}_kv^i+\gamma v^iv^k\partial_k\gamma) \right)\,.
\label{comp-heat-flux}
\end{flalign}
Since our assumptions imply $\partial_t{\cal T}=0$, $\bar{\nabla}_iv_j+\bar{\nabla}_jv_i=0$,
$v^i\partial_i\gamma=0$, $\partial_i\gamma=-\gamma^3v^k\bar{\nabla}_kv_i$, \eqref{comp-heat-flux}
boils down to
\begin{equation}
q^t = -\kappa\gamma^2 v^i\partial_i{\cal T}\,, \qquad
q^i = -\kappa{\cal T}\partial^i\ln\frac{\cal T}{\gamma}+v^iq^t\,.
\end{equation}
The vanishing of $q^\mu$ gives thus $\partial^i\ln({\cal T}/\gamma)=0$, i.e.,
${\cal T}/\gamma=\tau$, where $\tau$ is a constant.\\

{\bf Prop.~\ref{prop-diff-curr}:}

The components of the diffusion current in \eqref{subl-currents} read
\begin{equation}
q_{\text e}^t = -D\left((\gamma^2-1)\partial_t\frac{\mu}{\cal T}+\gamma^2 v^i\partial_i
\frac{\mu}{\cal T}\right)\,, \quad
q_{\text e}^i = -D\left(\gamma^2 v^i\partial_t\frac{\mu}{\cal T}+(\bar{g}^{ij}+\gamma^2
v^iv^j)\partial_j\frac{\mu}{\cal T}\right)\,. \label{comp-diff-curr}
\end{equation}
Stationarity implies $\partial_t(\mu/{\cal T})=0$, hence \eqref{comp-diff-curr} reduces to
\begin{equation}
q_{\text e}^t = -D\gamma^2 v^i\partial_i\frac{\mu}{\cal T}\,, \qquad
q_{\text e}^i = -D\partial^i\frac{\mu}{\cal T}+v^i q_{\text e}^t\,.
\end{equation}
$q_{\text e}^\mu=0$ leads thus to $\partial^i(\mu/{\cal T})=0$, i.e., $\mu/{\cal T}=\psi$, with
$\psi$ constant.

\end{document}